\documentclass[aps,prd,twocolumn,a4paper,superscriptaddress,10pt,floatfix]{revtex4-2}
\usepackage{graphicx,multirow,latexsym}
\usepackage{hyperref}
\usepackage[british]{babel}
\usepackage{amsmath,amsmath}

\begin{document}
\newcommand{\be}{\begin{equation}}
\newcommand{\ee}{\end{equation}}
\newcommand{\bq}{\begin{eqnarray}}
\newcommand{\eq}{\end{eqnarray}}
\newcommand{\bsq}{\begin{subequations}}
\newcommand{\esq}{\end{subequations}}
\newcommand{\bc}{\begin{center}}
\newcommand{\ec}{\end{center}}
\newcommand{\vos}[1]{{\left<#1\right>}}
\newcommand{\Hub}{\mathcal{H}}
\newcommand{\dd}[1]{\text{d}{#1}}
\renewcommand{\vec}[1]{\bf{#1}}
\newcommand{\matr}[1]{\boldsymbol{#1}}
\newcommand{\dpartial}[2]{\frac{\partial {#1}}{\partial {#2}}}
\newcommand{\ddfrac}[2]{\frac{\dd{#1}}{\dd{#2}}}
\newcommand{\ddpartial}[2]{\frac{\partial^2 {#1}}{\partial {#2}^2}}
\newcommand{\dddfrac}[2]{\frac{\dd{}^2 {#1}}{\dd {#2}^2}}
\newcommand{\dprime}[1]{{#1^\prime}}
\newcommand{\ddprime}[1]{{#1^{\prime\prime}}}

\title{Scaling solutions for current-carrying cosmic string networks}
\author{F. C. N. Q. Pimenta}
\email{up200908672@edu.fc.up.pt}
\affiliation{Centro de Astrof\'{\i}sica da Universidade do Porto, Rua das Estrelas, 4150-762 Porto, Portugal}
\author{C. J. A. P. Martins}
\email{Carlos.Martins@astro.up.pt}
\affiliation{Centro de Astrof\'{\i}sica da Universidade do Porto, Rua das Estrelas, 4150-762 Porto, Portugal}
\affiliation{Instituto de Astrof\'{\i}sica e Ci\^encias do Espa\c co, Universidade do Porto, Rua das Estrelas, 4150-762 Porto, Portugal}

\date{\today}

\begin{abstract}
Cosmic string networks are the best motivated relics of cosmological phase transitions, being unavoidable in many physically plausible extensions of the Standard Model. Most studies, including those providing constraints from and forecasts of their observational signals, rely on assumptions of featureless networks, neglecting the additional degrees of freedom on the string worldsheet, e.g. charges and currents, which are all but unavoidable in physically realistic models. An extension of the canonical velocity-dependent one-scale model, accounting for all such possible degrees of freedom, has been recently developed. Here we improve its physical interpretation by studying and classifying its possible asymptotic scaling solutions, and in particular how they are affected by the expansion of the Universe and the available energy loss or transfer mechanisms. We find three classes of solutions. For sufficiently fast expansion rates the charges and currents decay and one asymptotes to the Nambu-Goto case, while for slower expansion rates they can dominate the network dynamics. In between the two there is a third regime in which the network, including its charge and current, reaches full scaling. Under specific but plausible assumptions, this intermediate regime corresponds to the matter-dominated era. Our results agree with, and significantly extend, those of previous studies.
\end{abstract}
\maketitle
\section{\label{01intro}Introduction}

Cosmic strings are the most ubiquitous and interesting consequence of phase transitions in the early universe, arising naturally, though the Kibble mechanism \cite{Kibble76}, in both grand unified theories and superstring inspired inflation models. In the latter case, fundamental superstrings produced in the very early universe may have stretched to macroscopic scales, in which case they are known as cosmic superstrings \cite{Witten85}. Understanding their dynamics, evolution, and cosmological consequences, is an essential part of any credible attempt to understand fundamental cosmology \cite{VSbook,Copeland_2010}. While a fully quantitative study of cosmic superstrings is currently not possible, one can study general current-carrying string networks, which provide useful proxies for them. Such a study requires a combination of analytic modelling and numerical simulation (including robust statistical tools for post-processing these simulations); the present work addresses the former.

Analytic modelling of cosmic defect networks relies on developing a thermodynamics-type description which captures the dynamics of the system's relevant macroscopic quantities. This idea was first implemented by Kibble \cite{Kibble85}, who introduced a model of string networks with a single macroscopic correlation length. The current state of the art is the velocity-dependent one-scale (VOS) model \cite{VOS1,Martins_2002,Martins_2016}. It provides quantitative dynamical equations for both a characteristic length scale and the root mean square (RMS) velocity of the network, and has been fully calibrated against state-of-the-art field theory simulations \cite{Correia_2019,Correia_2023}.

A further degree of complexity stems from the fact that, in physically realistic models, one fully expects the string worldsheets to contain additional degrees of freedom, e.g. generic charges and currents, and even to be superconducting \cite{Witten_1984}. An extension of the VOS model, the charge-velocity-dependent one-scale model (CVOS), able to describe such generic charges and currents on the string worldsheet, has recently been developed \cite{Martins_2021,Rybak_2023}; previously developed models for the wiggly \cite{Vieira,Almeida_2021,Almeida_2022} and chiral \cite{Oliveira_2012} cases can be readily obtained as suitable limits of this general model. 

Both the VOS and CVOS models include several phenomenological free parameters, which ideally should be measured from numerical simulations, and this has been done for plain strings and also, to a lesser extent, for wiggly strings. However, at the time of writing this is not yet possible for generic current-carrying strings, despite steady progress towards this goal \cite{Saffin05,Urrestilla_2008,Lizarraga16,Correia_2022,Battye23}. In the present work, therefore, we study and classify all possible asymptotic scaling solutions of the CVOS model, in particular considering how they depend on the expansion rate of the universe and other physical mechanisms impacting the network dynamics. We will focus on the \emph{physical} scaling solutions, by which we mean those that could be cosmologically relevant, at least for some ranges (or specific values), of the model's phenomenological parameters. This is to be contrasted to other \emph{mathematical} solutions, which are asymptotic solutions of the CVOS equations but are not physically relevant, e.g. because they would imply a network of strings moving at the speed of light.

Our work follows in the footsteps of earlier analyses for the specific cases of chiral strings \cite{Oliveira_2012} and wiggly strings \cite{Almeida_2021,Almeida_2022}. Some solutions for the radiation and matter eras, and under the assumption of a linearised version of the CVOS model, have also been discussed in \cite{Linear}. Our present analysis is a generalisation of the earlier ones, recovering several of their results and finding analogous solutions in several cases (although our assumptions on the model's phenomenological parameters are sometimes different), but we also find new solutions, e.g. for non-chiral networks. Interestingly, we show that the fact that scaling solutions can be divided into three classes, primarily determined by the cosmological expansion rate, already noticed in the two particular cases, still applies in the general case. For a sufficiently damped network (due to a fast enough expansion rate) the charges and currents decay and the network evolves towards the plain Nambu-Goto case. Conversely, for sufficiently slow expansion rates the charge and current can dominate the network dynamics, preventing the linear scaling behaviour typical of plain networks. Last but not least, in between these two regimes, and for one specific expansion rate, there is a third regime in which the network is in full scaling---in other words, the canonical linear scaling regime also includes the charge and current. The exact value of this threshold expansion rate depends on the values of the model parameters, but under some assumptions it can occur at the matter-dominated era. Our results agree with, and significantly extend, those of previous studies.

The structure of the rest of the paper is as follows. In Sect. \ref{02VOS} we provide a brief introduction to the CVOS, model, mainly for the purposes of making the present work self-contained and introducing the relevant quantities, as well as describing the assumptions made in our analysis. Solutions in Minkowski spacetime are described in Sect. \ref{03nomechanisms}. Our main results are in Sects. \ref{03nolosses} and \ref{04losses}, which discuss the possible asymptotic scaling solutions in expanding universes, without and with losses respectively, and the conditions under which each of them may occur. A broader discussion of these solutions and our conclusions are then presented in Sect. \ref{05conclusions}, including a comparison with earlier results on chiral and wiggly strings. Finally, we note that in the present work we only address the case of unbiased networks (a term to be defined in the next section); the case of biased networks will be discussed in a follow-up work.

\section{\label{02VOS}The CVOS Model}

The canonical VOS model is clearly insufficient to accurately model the evolution of cosmic string networks possessing additional degrees of freedom on the string worldsheet, either in the form of small scale structure (commonly known as wiggles) or, as in our present case, in the form of more generic charges and currents. One of its physical assumptions is that a single length scale describes both the network correlation length and the network energy, but this can no longer hold in such situations: the additional degrees of freedom can themselves contribute to the energy balance \cite{Martins_2006}. The VOS equations must therefore be generalised.

A common approach is to keep the definition of the correlation length scale, $\xi$, as the typical string separation, and hence related to the so-called bare string energy, while defining an additional length scale, $L$, which relates to the total energy of the network. (In the case of featureless Nambu-Goto strings the two will coincide.) This has been explored in some detail for the wiggly case \cite{Vieira,Almeida_2022}, while the current carrying case has been much less explored, and only in the chiral limit \cite{Martins_1998}. More recently the CVOS model, able to self-consistently describe generic charges and currents, has been developed \cite{Martins_2021}. In what follows we provide a brief introduction to the model, focusing on the aspects that are directly relevant to our subsequent discussion. We refer the reader to the original work for more detailed derivations and also for a discussion of the model's underlying assumptions.

\subsection{CVOS dynamical equations}

To describe charges and currents on the string worldsheet one may start with the action  \citep{Martins_2021}
\begin{equation}
    S=-\mu_0\int f(\kappa)\sqrt{-\gamma}\dd\sigma^2
\end{equation}
where $\mu_0$ has units of mass squared, and the generic function $f(\kappa)$ depends on the so-called state parameter $\kappa$ defined from a scalar field, $\varphi$, as:
\begin{equation}
    \kappa= q^2-j^2
    = \frac{1}{a^2\dprime{\vec{x}}^2}\left(\varepsilon^2\dot{\varphi}^2-\dprime{\varphi}^2\right)
\end{equation}
where the microscopic charge $q^2$ and current $j^2$ have been implicitly defined, $a$ and $\varepsilon$ are the scale factor and coordinate energy per unit length respectively, and the last equality highlights the fact that the chiral limit studied in \cite{Oliveira_2012} corresponds to the limit $\kappa\to 0$. It is also convenient to introduce the dimensionless variables \cite{Rybak_2017}
\begin{subequations}
\bq
    \tilde{U}&\equiv & f-2q^2\frac{\dd f}{\dd \kappa} \\
    \tilde{T}&\equiv & f+2j^2\frac{\dd f}{\dd \kappa}\\
    \Phi&\equiv & -2qj\frac{\dd f}{\dd \kappa}\,
\eq
\end{subequations}
whose dimensional counterparts are the energy per unit length, the string tension, and the scalar field representing the current.

It follows that the total energy of the network is
\begin{equation}
     E= a\mu_0\int\tilde{U}\varepsilon\dd\sigma = a\mu_0\int f \varepsilon \dd\sigma -a\mu_0\int 2q^2\frac{\dd f}{\dd \kappa}\varepsilon\dd\sigma\,;
\end{equation}
the first term is the bare string energy as defined in the canonical VOS model (henceforth denoted $E_0$), which motivates the definition of the macroscopic ratio of the total and bare energies,
\begin{equation}
    \frac{E}{E_0} = \vos{f} - 2\vos{q^2\frac{\dd f}{\dd \kappa}} = F - 2Q^2\dprime{F}\,,
\end{equation}
where capital letters should be understood as the expected value of their microscopic counterparts and the variables have been assumed to be uncorrelated to obtain the last equality. Finally, one may rewrite the energy relation above in comoving units and also make explicit the relation between the comoving length scale, $L_c$, and its correlation length counterpart by defining
\begin{equation}
     \xi_c= L_c \sqrt{F-2Q^2\dprime{F}} = W L_c\,,
    \label{eq:lc_xic}
\end{equation}
where the last relation defines the convenient function $W$. Note that this function is the square root of the ratio of the total and bare energies, and therefore it is always larger than (or at most equal to) unity.

Under these assumptions, one can start with the microscopic equations of motion and derive the following averaged macroscopic equations \cite{Martins_2021}
\begin{subequations}
\label{eq:gvos}
\bq
    \dot{L}_c &=& \Hub L_c\left[v^2-\left(1-v^2\right)\frac{Q^2+J^2}{W^2}\dprime{F}\right] \\
    \dot{v} &=& \left(1-v^2\right)\left[\frac{k_v}{WL_c}\left(1+2\frac{Q^2+J^2}{W^2}\dprime{F}\right)\right]\nonumber\\
    & &-\left(1-v^2\right)\left[2v\Hub\left(1+\frac{Q^2+J^2}{W^2}\dprime{F}\right)\right]\\
    \left(J^2\right){\dot{}} &=& 2J^2\left(\frac{vk_v}{L_cW}-\Hub\right)\\
    \left(Q^2\right){\dot{}}  &=& 2Q^2\frac{\dprime{F}+2J^2\ddprime{F}}{\dprime{F}+2Q^2\ddprime{F}}\left(\frac{vk_v}{L_cW}-\Hub\right) \\
    \dot{\xi}_c &=& \Hub\xi_c v^2+\frac{Q^2+J^2}{W^2}\left(\Hub\xi_c v^2-v k_v\right)\dprime{F}\,;
\eq
\end{subequations}
here dots denote derivatives with respect to conformal time and $k_v$ is the usual momentum parameter \cite{Martins_2002}. Note that the last equation is not independent from the others.

However, these equations do not include additional energy loss mechanisms (apart from the cosmological expansion, which is accounted for by the comoving Hubble parameter $\Hub$). Such loss mechanisms include not only losses due to loop production but also possible charge and/or current losses (which are different nonlinear processes and can occur at different rates), and also bias parameters, describing e.g. whether a region with a high charge or current is more or less likely to be part of loop production events. Naturally these require phenomenological modelling, which is described in detail in \cite{Martins_2021,Linear,Rybak_2023}. Here we simply state the outcome of this analysis, which is the addition of the following terms to the previous equations
 \begin{subequations}
\bq
    \dot{L}_c &=& \dots + \frac{g}{W}\frac{\tilde{c}}{2}v\\
    \left(J^2\right){\dot{}} &=& \dots + \rho\tilde{c}\frac{v}{L_c}\frac{(g-1)W}{\dprime{F}-2Q^2\ddprime{F}}\\
    \left(Q^2\right){\dot{}} &=&\dots + (1-\rho)\tilde{c}\frac{v}{L_c}\frac{(g-1)W}{\dprime{F}+2Q^2\ddprime{F}}\\
    \dot{\xi}_c &=& \dots + \frac{\tilde{c}}{2}v\,,
\eq
\end{subequations}
while the velocity equation is unchanged. Here $\tilde{c}$ is the usual loop chopping efficiency (which already exists in the canonical VOS model, cf. \cite{VOS1}) and is expected to be a constant, while $g$ and $\rho$ are bias parameters and are not necessarily constants. The first of these encodes possible overall biases of the additional degrees of freedom with respect to the bare string, and should be $g=1$ in the unbiased case. The second parameter, which is only relevant if $g\neq1$, describes biases between charge and current, and should be $\rho=1/2$ in the unbiased case.

Regarding the $g$ bias function, a detailed analysis in \cite{Rybak_2023} suggests that it is expected to have the following generic form
\begin{equation}
    g = 1- g_Q\frac{\dprime{F}+2Q^2\ddprime{F}}{F-2Q^2\dprime{F}}Q^2-g_J\frac{\dprime{F}-2Q^2\ddprime{F}}{F-2Q^2\dprime{F}}J^2\,,
\end{equation}
where $g_Q$ and $g_J$ are dimensionless constants describing how much time-like and space-like components of the network's current are lost due to loops production.

For future reference we note that this form makes it clear that solutions with decaying charge and current will asymptotically have $g=1$, while for constant charge and/or current solutions one may still expect constant, but not necessarily unity, values for $g$. Finally, its asymptotic value for growing charges and/or currents depends on the nature of the microscopic model. Since in these cases, as we show in what follows, we must have $\dprime{F}\neq 0$, one may distinguish between models where $\ddprime{F}=0$ and models where $\ddprime{F}\neq0$. In the first case, it follows that
\begin{equation}
    g = 1 + \frac{g_Q}{2}+\frac{g_J}{2}\frac{J^2}{Q^2}\,,
\end{equation}
implying that if both charge and current exhibit the same behaviour, this is still a constant. On the other hand, if $\ddprime{F}\neq 0$, it follows that:
\begin{equation}
    g = 1 + (g_QQ^2-g_JJ^2)\frac{\ddprime{F}}{\dprime{F}}\,,
\end{equation}
and now the bias function exhibits a dependency on the charge and current values, which would only vanish if they both exhibit similar behaviour and
\begin{equation}
    \frac{g_Q}{g_J}=\frac{J^2}{Q^2}\,,
    \label{eq:growing_cond}
\end{equation}
in which case we find $g=1$ once again. In any other case, $g$ will have an implicit time dependency. In the following sections of the present work we only discuss unbiased solutions, i.e. those with $g=1$; the biased solutions will be discussed in a subsequent work.

\subsection{Scaling solutions}

The CVOS model equations provide an analytic and quantitative macroscopic description of the network evolution, but they are still a formidable problem to solve analytically and further simplifications or assumptions are needed in order to gain useful physical insight on the model behaviour. An alternative would be to explore the model numerically, and some examples of this can be found in \cite{Linear,Rybak_2023}. We leave such an exploration for future work, and in what follows our approach is to look for asymptotic scaling solutions of power law form. Specifically, we define
\begin{subequations}
\label{powerlaws}
\bq
    L_c &=& L_0\tau^\alpha \\
    v &=& v_0\tau^\beta \\
    J^2 &=& J_0^2 \tau^\gamma \\
    Q^2 &=& Q_0^2 \tau^\delta \\
    \xi_c &=& \xi_0 \tau^\varepsilon  \\
    W &=& W_0\tau^\zeta = \frac{\xi_0}{L_0}\tau^{\varepsilon-\alpha}\,,
\eq
\end{subequations}
where the physical solutions must be such that $\beta\leq 0$ and $\alpha\leq 1$. Clearly this is a simplifying assumption, although it is amply justified by previous experience of analytic modelling (and also of high-resolution numerical simulation) of cosmic string networks. It is especially so if the scale factor is also assumed to have a power law form. In terms of conformal time, $\tau$,
\begin{equation}
    a = a_0 \tau^{\lambda}\,,\qquad \mathcal{H} = \frac{\dprime{a}}{a} = \frac{\lambda}{\tau}\,,
\end{equation}
or equivalently in terms of physical time, $t$,
\begin{equation}
    a(t) \propto t^{\frac{\lambda}{1+\lambda}}\,.
\end{equation}
This assumption is manifestly convenient numerically, but also reasonable in the real universe, e.g. $\lambda=1$ corresponds to the radiation era and $\lambda=2$ to the matter era.

It should be clear that this description in terms of conformal time and comoving length scales can be easily reinterpreted as a function of cosmic time and physical lengths. Any quantity that is scaling with respect to conformal time as a power law of exponent $\zeta$, will scale as a function of cosmic time as a power law with exponent $\zeta/(1+\lambda)$, while if a comoving length scales as $\tau^\kappa$ its physical counterpart will scale as $\tau^{\kappa+\lambda}$. In particular, and combining both of these results, it should be noted that linearly scaling comoving distances with respect to conformal time are fully equivalent to linearly scaling physical quantities with respect to cosmic time. In other words, the presence or absence of the usual linear scaling solution will be clear in either description.

By assuming these power law solutions, the CVOS equations take the following generic form
\begin{subequations}
\bq
    \alpha  &=&  \lambda \left[v_0^2\tau^{2\beta}-C_v\mathcal{K}\tau^{2\left(\alpha-\varepsilon\right)+\eta}\right] + \frac{g\tilde{c}}{2}\frac{v_0}{\xi_0} \tau^{1+\beta-\varepsilon} \\
    \beta &=& C_v\left[\frac{k_v}{v_0\xi_0}\tau^{1-\beta-\varepsilon}\left(1+2\mathcal{K}\tau^{2\left(\alpha-\varepsilon\right)+\eta}\right)\right]\nonumber\\
    && -C_v\left[2\lambda \left(1+\mathcal{K}\tau^{2\left(\alpha-\varepsilon\right)+\eta}\right)\right]\\
    \gamma &=& 2\left(\frac{v_0k_v}{\xi_0}\tau^{1+\beta-\varepsilon}-\lambda\right)\nonumber\\
    && -\rho\tilde{c}\frac{v_0}{\xi_0}\frac{\xi_0^2}{J_0^2L_0^2}\frac{1-g}{\dprime{F}-2Q_0^2\tau^\delta\ddprime{F}}\tau^{1+\beta+\varepsilon-2\alpha-\gamma}
    \\
    \delta  &=& 2\frac{\dprime{F}+2J_0^2 \tau^\gamma\ddprime{F}}{\dprime{F}+2Q_0^2 \tau^\delta\ddprime{F}}\left(\frac{v_0k_v}{\xi_0}\tau^{1+\beta-\varepsilon}-\lambda\right) 
    -\nonumber \\
    && -(1-\rho)\tilde{c}\frac{v_0}{\xi_0}\frac{\xi_0^2}{Q_0^2L_0^2}\frac{1-g}{\dprime{F}+2Q_0^2\tau^\delta\ddprime{F}}\tau^{1+\beta+\varepsilon-2\alpha-\delta}
    \\
    \varepsilon &=&\lambda v_0^2\tau^{2\beta}\left(1+\mathcal{K}\tau^{2\left(\alpha-\varepsilon\right)+\eta}\right)\nonumber\\
    && - \frac{v_0k_v}{\xi_0}\mathcal{K}\tau^{2\left(\alpha-\varepsilon\right)+\eta+1+\beta-\varepsilon}+\frac{\tilde{c}}{2}\frac{v_0}{\xi_0}\tau^{1+\beta-\varepsilon}
\eq
\end{subequations}
where for subsequent convenience the following additional quantities have been defined
\begin{subequations}
\bq
    \mathcal{K} &=& \frac{L_0^2\mathcal{J}_0^2}{\xi_0^2}\dprime{F}\\
    \mathcal{J}_0^2 &=&
    \begin{cases}
        Q_0^2+J_0^2&\text{, if }\delta=\gamma\\
        J_0^2&\text{, if }\delta<\gamma\\
        Q_0^2&\text{, if }\delta>\gamma \\
    \end{cases}\\
    \eta &=&
    \begin{cases}
    \gamma & \text{, if }\delta\leq\gamma\\
    \delta& \text{, if }\delta>\gamma\\
    \end{cases}\\
    C_v &=& 1-v^2\,.
\eq
\end{subequations}  
Note that the last of these definitions is reasonable since for $\beta=0$, $1-v_0^2$ is a constant, and for $\beta<0$, $1-v_0^2t^{2\beta}\sim1$, which is also constant. Along the same lines, we note that the momentum parameter $k_v$ is velocity-dependent \cite{Martins_2002}, but it will also be a constant in both of these cases. Finally, it has been assumed that all the pre-factor parameters in Eq.(\ref{powerlaws}) are non-zero. 

A preliminary inspection of the different equations readily identifies some of the parameters which play a central role in the analysis to come. Most notably, the factor $2(\alpha-\varepsilon)+\eta$ appears associated with $\dprime{F}$ and is closely related to Eq.(\ref{eq:lc_xic}), which in terms of the power law defined above yields
\begin{equation}
    \frac{\xi_0^2}{L_0^2}\tau^{2(\varepsilon-\alpha)}=F-2Q_0^2\tau^\delta\dprime{F}\,.
\end{equation}
This is relevant since the presence of $F$ (which is not expected to vanish) on the right hand side provides a constraint on the relation between the three exponents. In particular, for decaying or constant charge solutions ($\delta\leq0$) both length scales must evolve with a similar rate, $\alpha=\varepsilon$. This is of course expected, as in that case the ratio of total and bare string energies should be a constant (which will be unity in the Nambu-Goto limit). On the other hand, growing charge cases are only possible if the overall characteristic length grows slower than the correlation length, again as it should be. In any case, it is not possible to have solutions where $\alpha>\varepsilon$, unless a rather unphysical behaviour is prescribed where $F=0$.

It should be mentioned that one may find various mathematical solutions to the evolution equations that are not physical for different reasons. Specifically, we can straightforwardly identify three such physical restrictions.  Firstly, the correlation length cannot scale faster than $t$ (or the comoving correlation length faster than $\tau$), since physically that would violate causality. (At a phenomenological level, that could be interpreted as the network decaying and disappearing.) Secondly, the network RMS velocity must be smaller than the speed of light, which in our units corresponds to $v<1$. And thirdly, in an expanding universe, the momentum parameter is expected to be non-zero---although, conversely, it is expected to vanish in Minkowski spacetime. Our analysis, reported in the following sections, will focus on the physical solutions, but we will also briefly discuss the Minkowski spacetime as a starting point. Solutions with $v=1$ can be found in \cite{Thesis}.

\section{\label{03nomechanisms}Solutions without expansion}

The simplest case which can be analysed leads to the solutions in Minkowski spacetime. In this branch of solutions, one may set the momentum parameter and the expansion rate to null values ($k_v=0$, $\lambda=0$), reducing the equations to:
\begin{subequations}
\bq
    \alpha  &=&  \varepsilon = \frac{\tilde{c}}{2}\frac{v_0}{\xi_0} \tau^{1+\beta-\varepsilon} \\
    \beta  &=& \gamma = \delta = 0
\eq
\end{subequations}
It should be clear that without any loss mechanism ($\tilde{c}=0$), every single quantity of interest will remaining constant and the only possible solution is given by
\begin{subequations}
\bq
    L_c &=& L_0\\
    \xi_c&=&\xi_0 \\
    J^2 &=& J_0^2\\
    Q^2 &=& Q_0^2 \\
    v &=& v_0\,,
\eq
\end{subequations}
which is a manifestation of energy conservation.

On the other hand, allowing for non vanishing energy loss parameters, one finds the solution
\begin{subequations}
\bq
    L_c &=& L_0\tau\\
    \xi_c&=&\xi_0\tau \\
    J^2 &=& J_0^2\\
    Q^2 &=& Q_0^2 \\
    v &=& v_0\,,
\eq
\end{subequations}
subject to the constraint
\bq
    \tilde{c}&=&\frac{2\xi_0}{v_0}\,.
\eq
We are not aware of any reliable Minkowski space simulations of current-carrying string networks (which may be more challenging than expanding universe ones, given the absence of mechanisms damping the radiation in the simulation box), but it would be particularly interesting to numerically confirm this solution.

\section{\label{03nolosses}Cosmological solutions without energy loss mechanisms}

We start once more by considering the case without energy losses (other than the obvious one due to the cosmological expansion), which obtains by simply setting $\tilde{c}=0$. While this assumption is not expected to be fully realistic, it has the advantage of providing initial insight into the possible scaling solutions (with considerably simpler CVOS equations), which also provide a benchmark against which the solutions with energy losses (to be addressed in the next section) can be compared. Indeed, it will be seen that the general solutions are all extensions of the ones discussed herein.

From the equation for the current one can see that in the absence of energy loss mechanisms any physical solution in an expanding universe must be such that $\varepsilon\geq 1+\beta$. If one assumes $\varepsilon>1+\beta$, then only decaying current solutions are possible, but further exploration quickly shows that there are no consistent solutions in such a branch (i.e., solutions which would satisfy all equations). On the other hand, if the sub-branch with $\varepsilon=1+\beta\leq1$ is assumed, the CVOS equations simplify to
\begin{subequations}
\bq
    \alpha  &=&  \lambda \left[v_0^2\tau^{2\beta}-C\mathcal{K}\tau^{2\left(\alpha-\varepsilon\right)+\eta}\right]  \\
    \beta &=& \frac{C_vk_v}{v_0\xi_0}\tau^{-2\beta}\left(1+2\mathcal{K}\tau^{2\left(\alpha-\varepsilon\right)+\eta}\right)\nonumber\\
    && -2C_v\lambda\left(1+\mathcal{K}\tau^{2\left(\alpha-\varepsilon\right)+\eta}\right)\\
    \gamma &=& 2\left(\frac{v_0k_v}{\xi_0}-\lambda\right)
    \\
    \delta  &=& 2\frac{\dprime{F}+2J_0^2 \tau^\gamma\ddprime{F}}{\dprime{F}+2Q_0^2 \tau^\delta\ddprime{F}}\left(\frac{v_0k_v}{\xi_0}-\lambda\right) 
    \\
    \varepsilon &=& \lambda v_0^2\tau^{2\beta}\left(1+\mathcal{K}\tau^{2\left(\alpha-\varepsilon\right)+\eta}\right)\nonumber\\
    && -\frac{v_0k_v}{\xi_0}\mathcal{K}\tau^{2\left(\alpha-\varepsilon\right)+\eta}
\eq
\end{subequations}
From a physical point of view, we note that the linear bare energy scaling with constant velocity that constitutes the canonical Nambu-Goto network evolution is a possible solution---recall that this solution can occur, for some expansion rates, even without loop production \cite{Martins_2016,Almeida_2021}.

It is also clear that under the assumptions of this sub-branch the quantity $\mathcal{K}\tau^{2\left(\alpha-\varepsilon\right)+\eta}$ plays a central role. In particular, its value is completely fixed at $-1/2$ for decaying velocity solutions. In that case, the only consistent solution, for an expansion rate of $\lambda=2/3$, is
\bsq
\label{solutionA1}
\bq
    L_c &=& L_0\tau^{1/3}\\
    \xi_c &=& \xi_0\tau^{1/3} \\
    J^2 &=& J_0^2\\
    Q^2 &=& Q_0^2\\
    v &=& v_0\tau^{-1/3}\,,
\eq
\esq
subject to the constraints
\bsq
\label{constraintsA1}
\bq
\mathcal{K}&=&-\frac{1}{2}\\
\lambda &=& \frac{v_0k_v}{\xi_0}=\frac{2}{3}\,.
\eq
\esq
Although this charge/current dominated solution is obtained here for a single expansion rate (which in physical time is $a\propto t^{2/5}$, slower than in the radiation era), it is actually compatible with the slow expansion rate branch identified by \cite{Oliveira_2012}, in the particular case where their phenomenological parameter $s$ is set to $s=0$. Although this may not be clear at first sight, it should be noted that for $s=0$ one of the constraints of the work by \cite{Oliveira_2012} fixes the charge value at $Q_0=1$ and the velocity equation is asymptotically reduced to
\begin{equation}
    \dot{v} = -\frac{1-v^2}{2}\left(2Hv\right) \approx -Hv 
\end{equation}
implying that, for power law solutions, one has
\begin{equation}
    v_0\beta t^{\beta-1} = -\lambda v_0 t^{\beta-1}
    \longrightarrow
    \lambda = -\beta
\end{equation}
While this relation is similar to the one just identified, although expressed with respect to the cosmic time, it should be further noted that another constraint identified by \cite{Oliveira_2012} is $\beta=1-\alpha=1-3\lambda/2$, which when combined with the condition above yields
\begin{equation}
    \lambda = \frac{3\lambda}{2} -1 \longrightarrow \lambda = \frac{2}{5}
\end{equation}
and a single expansion rate is allowed. Finally, one notes that an expansion rate of $2/5$ when expressed with respect to cosmic time is equivalent to an expansion rate of
\begin{equation}
    \lambda_t = \frac{\lambda_\tau}{\lambda_\tau+1} = \frac{2}{5}\longrightarrow
    \lambda_\tau = \frac{2}{3}
\end{equation}
which is exactly the same as in Eq. \ref{constraintsA1}. It is interesting to note that, in clear contrast with the results from \cite{Oliveira_2012}, there seems to be no solution for a wide range of expansion rates below the critical value. Since in \cite{Oliveira_2012} the term parameterised by $s$ is a cross-term, this may suggest that our assumption of a fully separable model has an impact on the allowed solutions. Finally, note that the earlier work only addressed the chiral limit, but in our solution above one does not require that $Q_0^2=J_0^2$; instead it is required that
\begin{equation}
    J_0^2 = -\frac{F}{2\dprime{F}}\,;
\end{equation}
to the extent that $F$ and $\dprime{F}$ encode information on the microphysics of the specific particle physics model under consideration (recall that they are macroscopic averages of microscopic quantities), this is an interesting expression, because it directly relates the allowed value of the scaling current with the model's microphysics.

Let us now consider the sub-branch of solutions with a constant velocity. Firstly, it should be noted that all such solutions must have a comoving correlation length which scales linearly with conformal time (or equivalently, as mentioned before, a physical correlation length which scales linearly with physical time). Here there are four possible solutions to consider.

One such solution, valid for expansion rates such that $v_0^2=1/\lambda$, is found in the particular case where $\dprime{F}=0$ and is given by
\bsq
\label{solutionA2}
\bq
    L_c &=& L_0\tau\\
    \xi_c &=&\xi_0\tau \\
    J^2 &=& J_0^2\tau^{4-2\lambda}\\
    Q^2 &=& Q_0^2\tau^{4-2\lambda} \\
    v &=& v_0\,,
\eq
\esq
subject to the constraints
\bsq
\label{constraintsA2}
\bq
    \dprime{F}&=&0\\
    \lambda &=& \frac{1}{v_0^2} >1\\
    \frac{v_0k_v}{\xi_0}&=&2 \\
    Q_0^2&=&J_0^2\,,
\eq
\esq
with the caveat that that last of the constraints is not applicable for $\lambda=2$, or equivalently $\gamma=\delta=0$. It should be noted that the above constraints imply that there is a strict lower bound for $\lambda$, specifically $\lambda>1$, and hence this solution is not possible in the radiation era, but it would be possible in the matter era---and in this particular case there is no requirement of equal charge and current. For expansion rates between radiation and matter domination, $1<\lambda<2$, one would have a growing charge and current. However, a bound $\lambda\ge2$ is physically more plausible, since it ensures $v_0^2\le1/2$. In such a case the matter era is the slowest allowed possible expansion rate, and exhibits a fully scaling (but not necessarily chiral) network, while for any larger ones we will have a decaying change and current, and the network evolves towards the featureless Nambu-Goto case.

Still within the sub-branch where $\dprime{F}=0$, there is another possible solution which can only occur for fast expansion rates,
\bsq
\label{solutionA3}
\bq
    L_c &=& L_0\tau\\
    \xi_c &=& \xi_0\tau \\
    J^2 &=& J_0^2\tau^{4-2\lambda}\\
    Q^2 &=& Q_0^2 \\
    v &=& v_0
\eq
\esq
subject to the constraints
\bsq
\label{constraintsA3}
\bq
\dprime{F}&=&0\\
\lambda &=& \frac{1}{v_0^2} >2\\
\frac{v_0k_v}{\xi_0}&=&2\,.
\eq
\esq
This solution therefore can't occur in the radiation or the matter era, so it is of limited cosmological significance, but it could be tested in numerical simulations. Moreover, the fact that the current decays while the charge is allowed to persist is somewhat counter-intuitive. Physically one expects that currents should be less sensitive to the damping caused by expansion than charges, and therefore one could also expect that electric and magnetic strings may occur for slow and fast expansion rates respectively---and indeed there is some support for this expectation coming from numerical simulations.

\begin{figure*}
    \centering
    \includegraphics[width=0.32\textwidth]{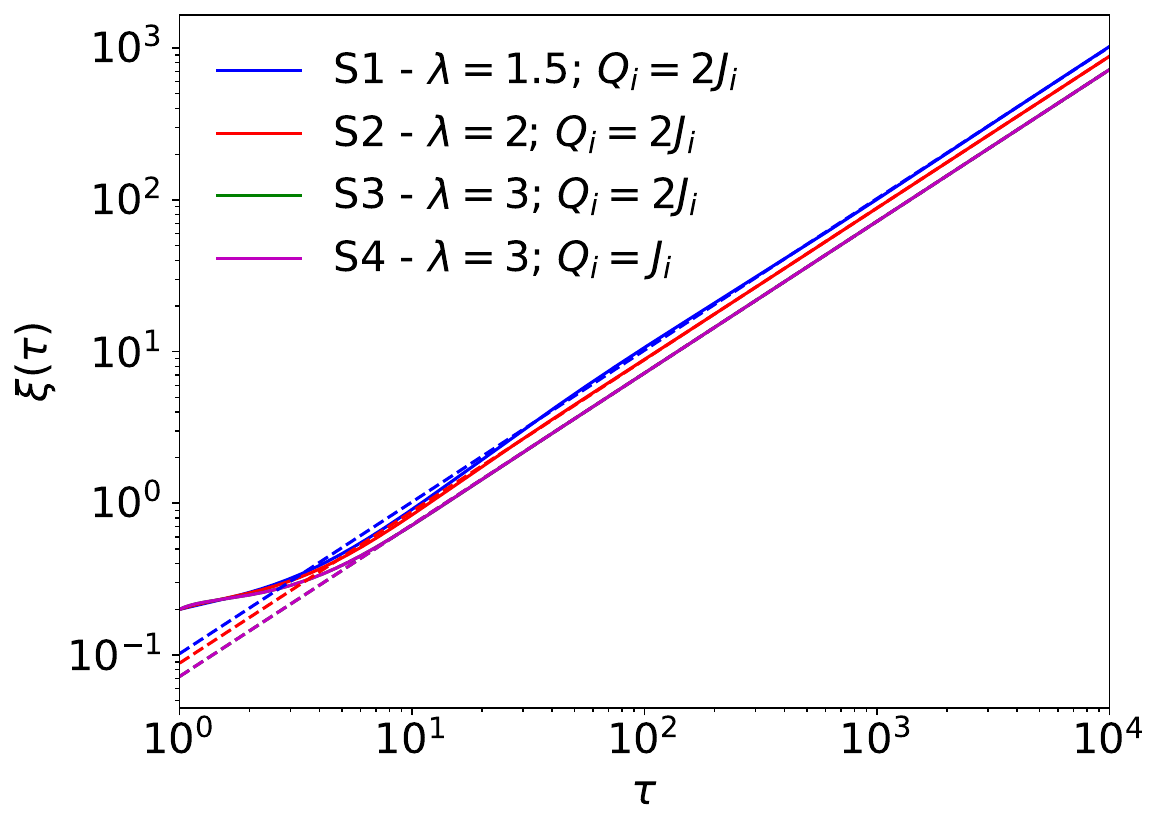}
    \includegraphics[width=0.32\textwidth]{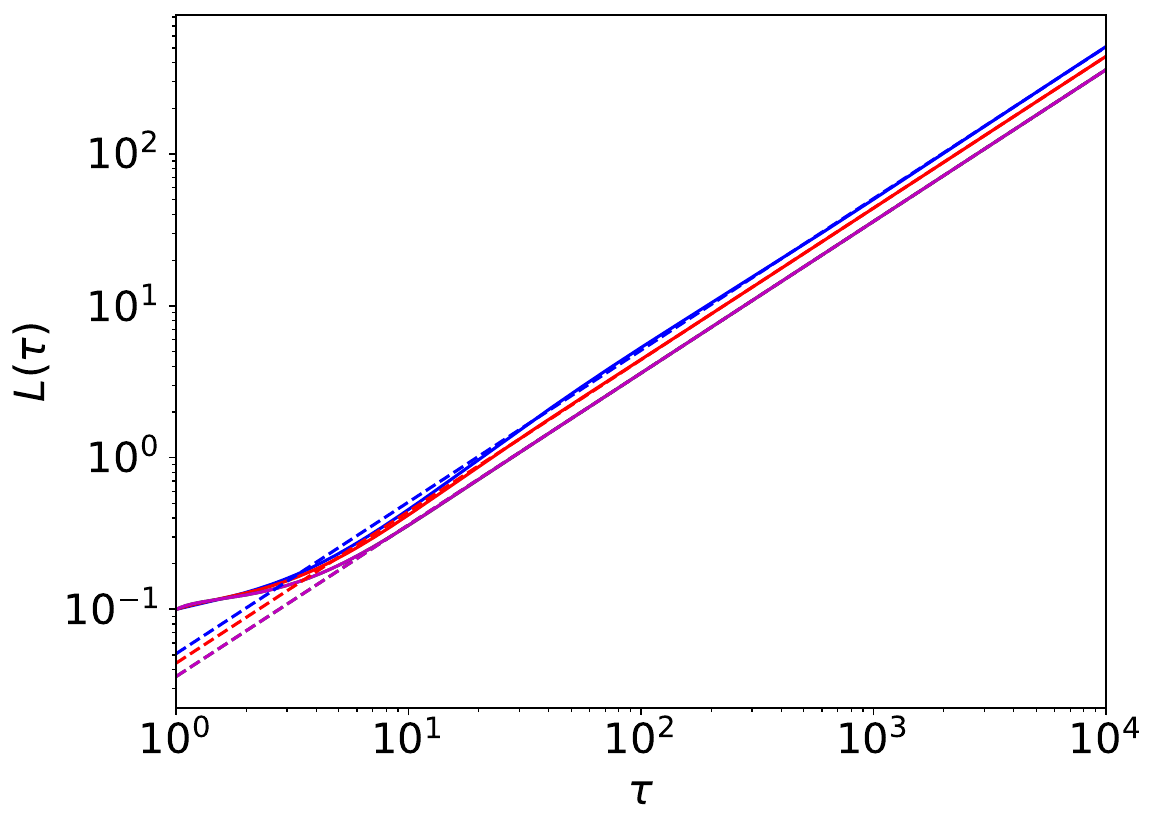}
    \includegraphics[width=0.32\textwidth]{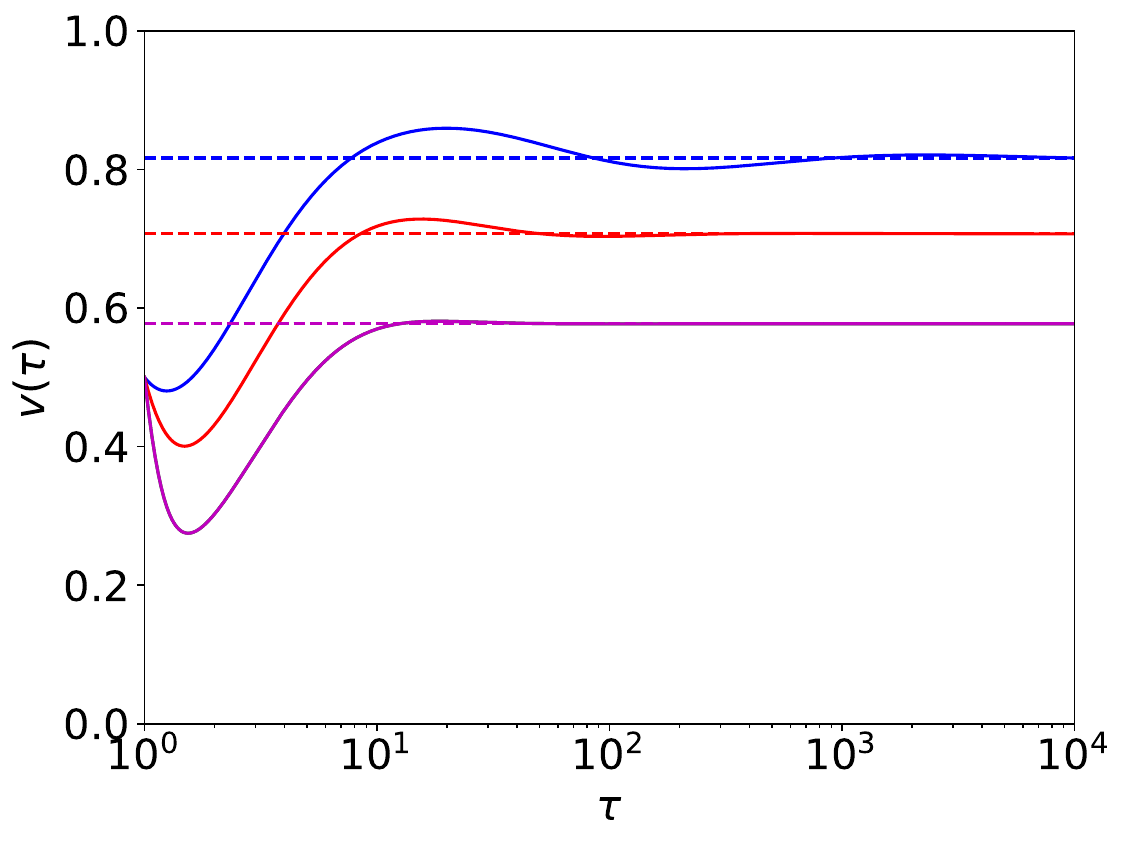}
    \includegraphics[width=0.32\textwidth]{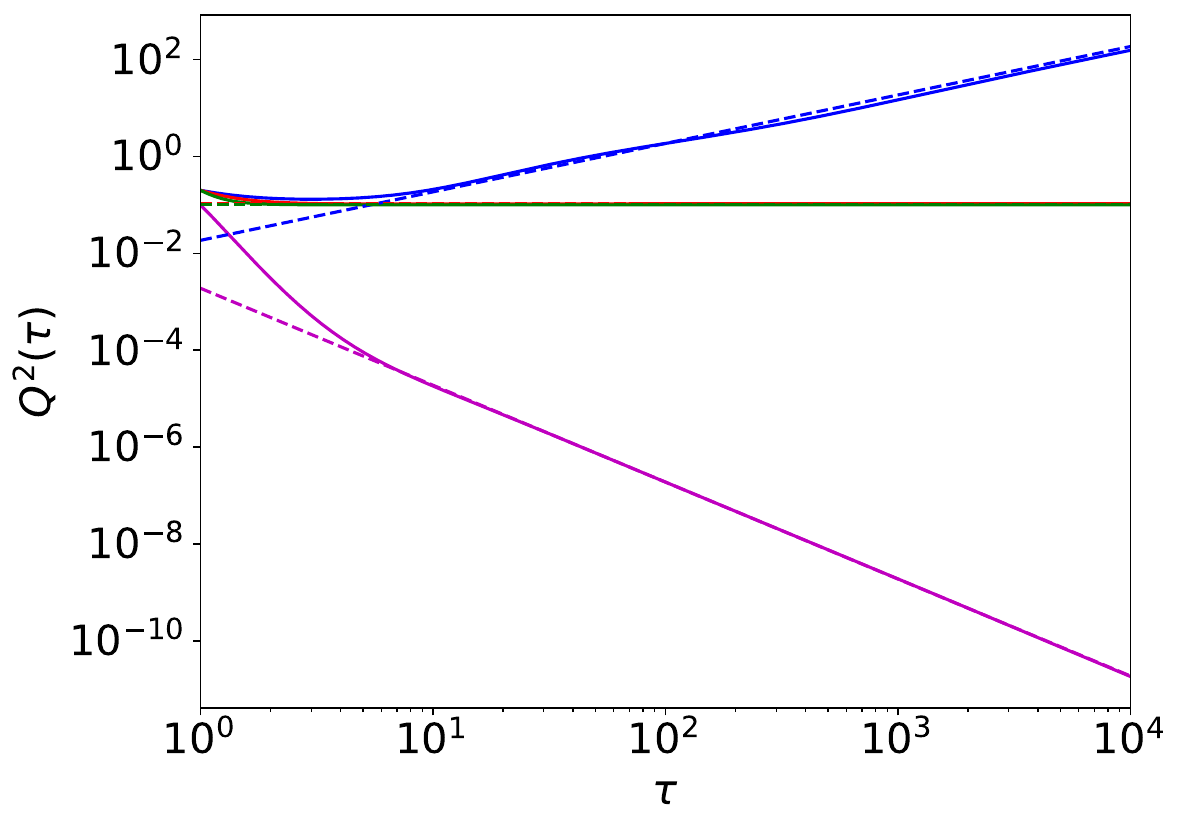}
    \includegraphics[width=0.32\textwidth]{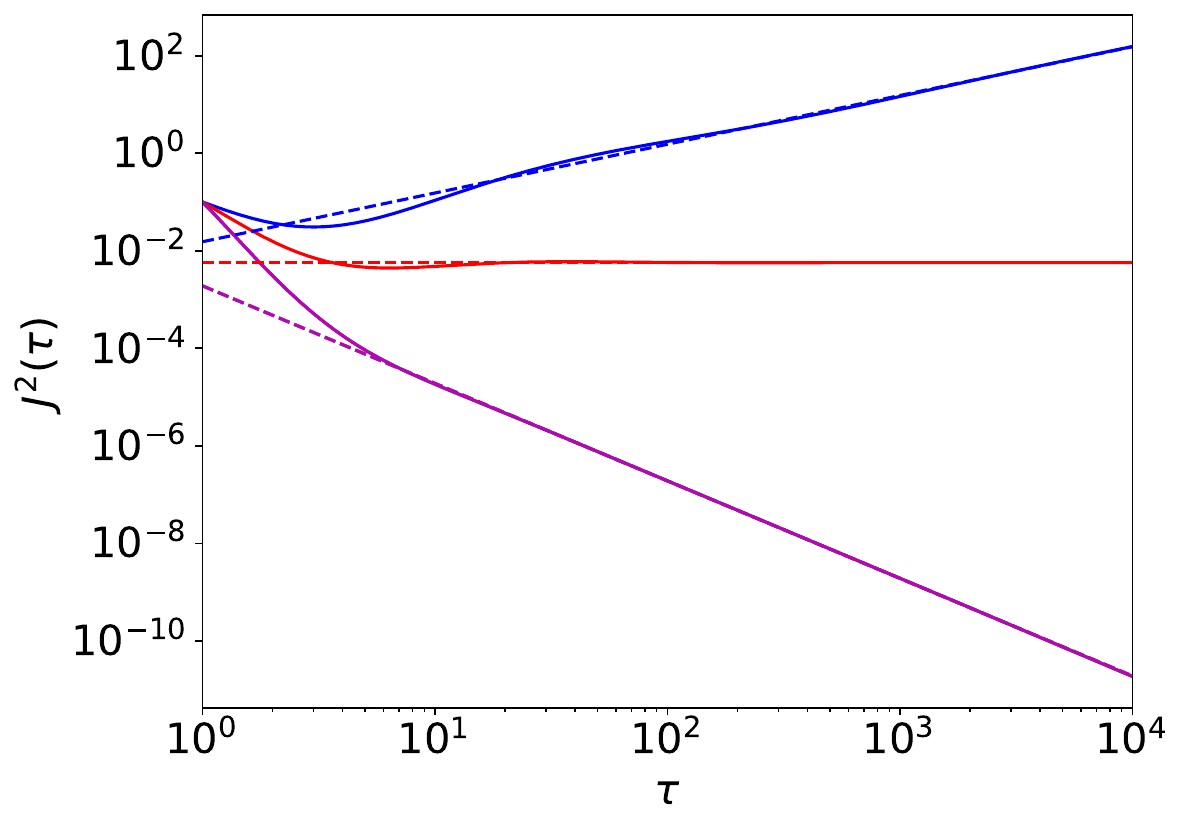}
    \includegraphics[width=0.32\textwidth]{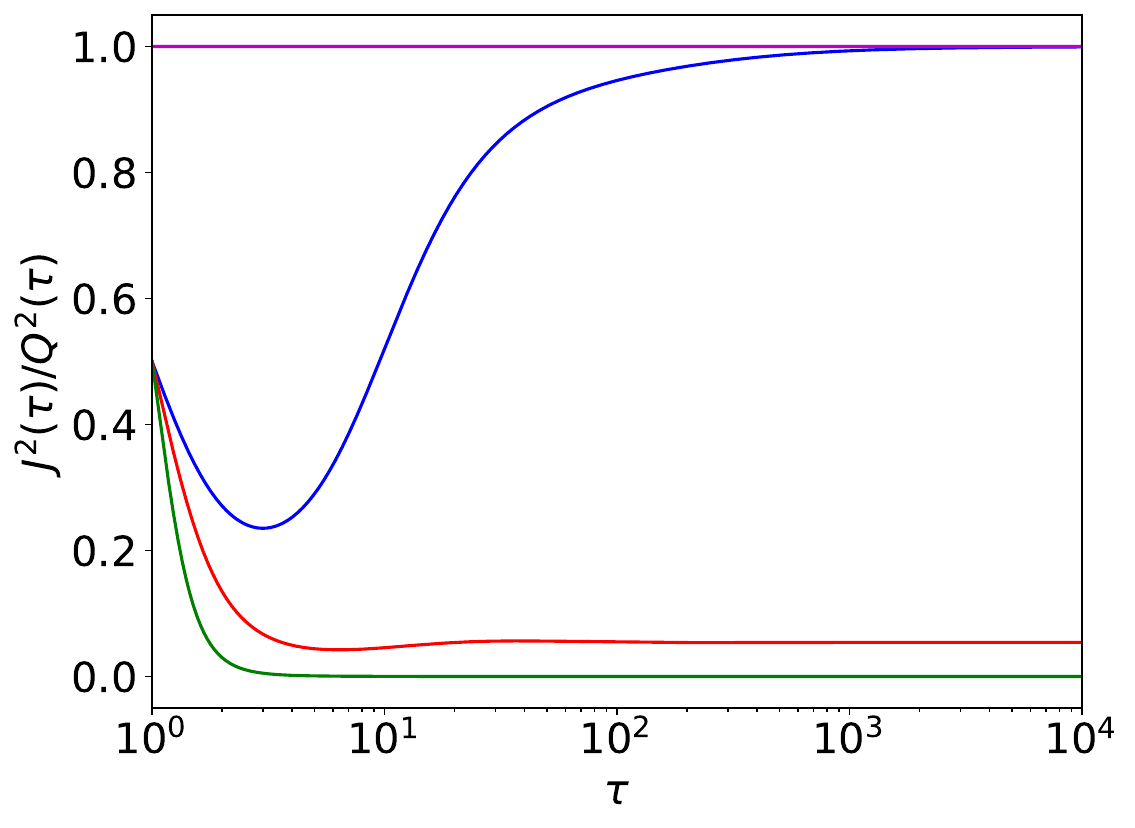}
    \caption{Network evolution as obtained by numerically solving the CVOS equations without loss mechanisms and assuming $\dprime{F}=0$, for different initial conditions and expansion rates. The four cases correspond to the following initial conditions: (Blue) $\lambda=3/2$, $Q_i=2J_i=0.2$; (Red) $\lambda=2$, $Q_i=2J_i=0.2$; (Green) $\lambda=3$, $Q_i=2J_i=0.2$; (Purple) $\lambda=3$, $Q_i=J_i=0.1$.}
    \label{fig01}
\end{figure*}

Figure \ref{fig01} depicts the network evolution for different expansion rates and $\dprime{F}=0$, showing that the above asymptotic solutions can be dynamically obtained from the CVOS model equations. In the first three of the simulated cases (shown in blue, red and green) the initial network charge and current were taken to be 0.2 and 0.1, respectively, and the constant charge solution is found. In the fourth (purple) case the initial charge and current were both assumed to be 0.1, and in this case the network evolves towards the decaying charge and current sub-branch. It should be noted that although the first (blue) case started with a charge to current ratio of 2, the asymptotic solution (shown with a dashed line) has equal charge and current, while the same does not happen for the second (red) case. This can be easily understood by noting that the growing current sub-branch is only possible if $Q_0=J_0$, and so the network finds its way into this state. Last but not least, we remark that the blue case, with $\lambda=3/2$, has an asymptotic velocity $v_0^2>1/2$, confirming our previous statement that a bound $\lambda\ge2$ is physically more plausible in this branch of solutions,

Leaving the cases where $\dprime{F}=0$ and assuming instead that $\dprime{F}\neq0$, one can find two additional solutions. Firstly, there is one which is very similar to that given by Eq.(\ref{solutionA2}), having the form
\bsq
\label{solutionA4}
\bq
    L_c &=& L_0\tau\\
    \xi_c&=&\xi_0\tau\\ 
    J^2 &=& J_0^2\tau^{4-2\lambda}\\
    Q^2 &=& Q_0^2\tau^{4-2\lambda} \\
    v &=& v_0\,,
\eq
\esq
subject to the constraints
\bsq
\label{constraintsA4}
\bq
    \lambda &=& \frac{1}{v_0^2} >2\\
    \frac{v_0k_v}{\xi_0}&=&2 \,.
\eq
\esq
As in the previous solution, given by Eqs. (\ref{solutionA3}--\ref{constraintsA3}), this asymptotically Nambu-Goto one is not possible in the radiation or matter eras. Here, however, the chiral condition does not have to hold and arbitrary current and charge values are possible. Comparing both solutions, it can be seen that fast expansion rates associated with decaying charges and currents are asymptotically equivalent to the canonical Nambu-Goto case, but only the chiral limit is allowed when $\dprime{F}=0$. Again. this highlights the relevance of the microphysics of the network in its cosmological dynamics.

\begin{figure*}
    \centering
    \includegraphics[width=0.32\textwidth]{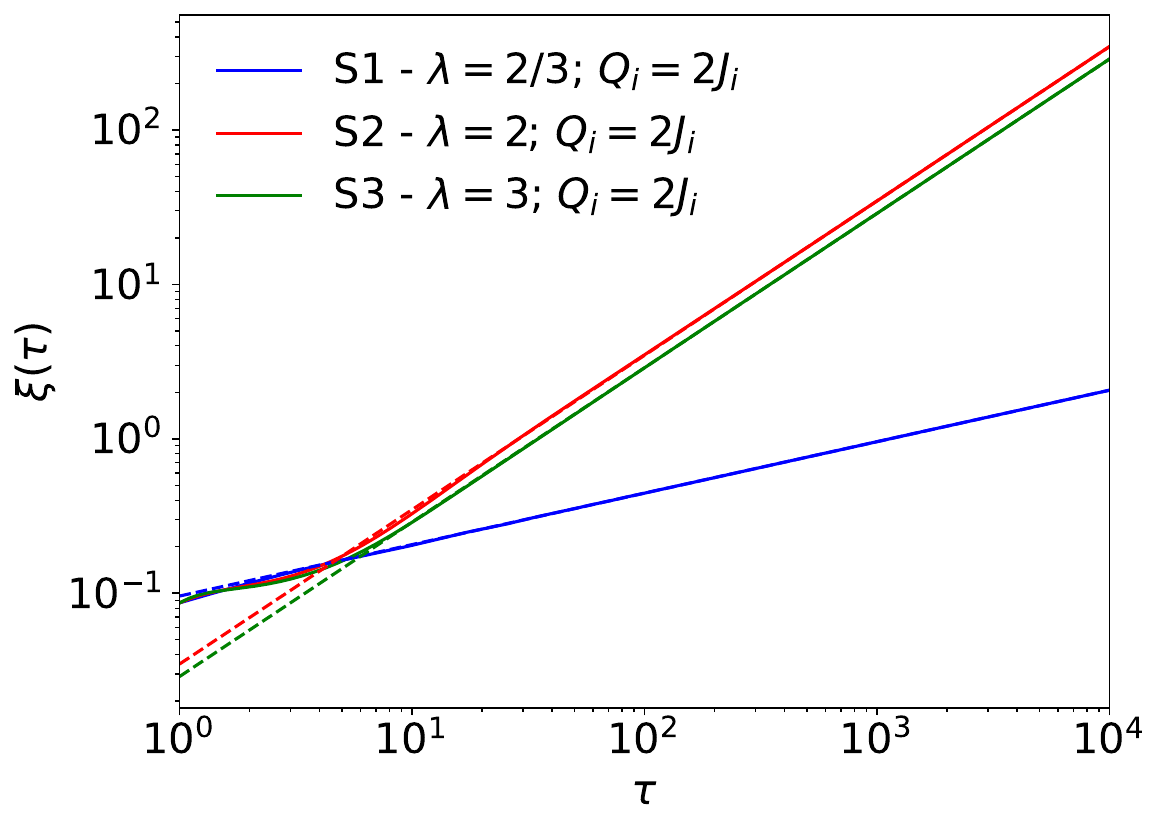}
    \includegraphics[width=0.32\textwidth]{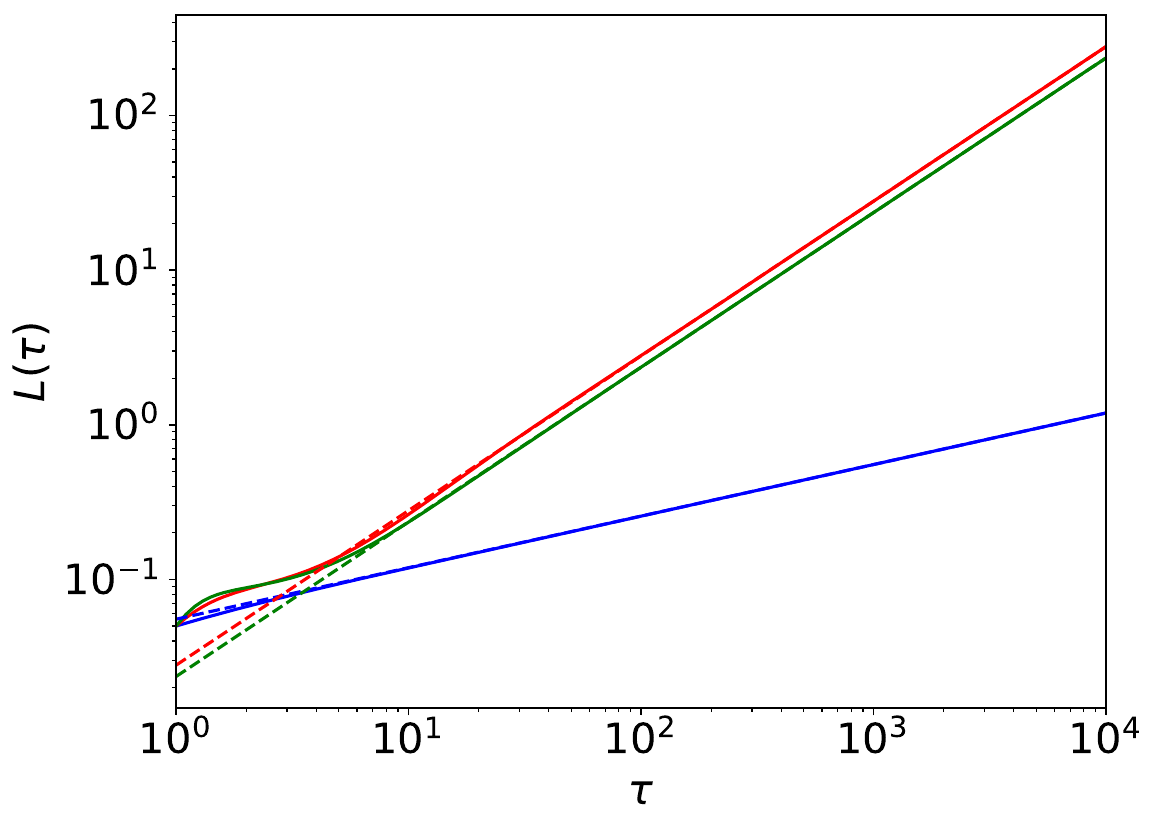}
    \includegraphics[width=0.32\textwidth]{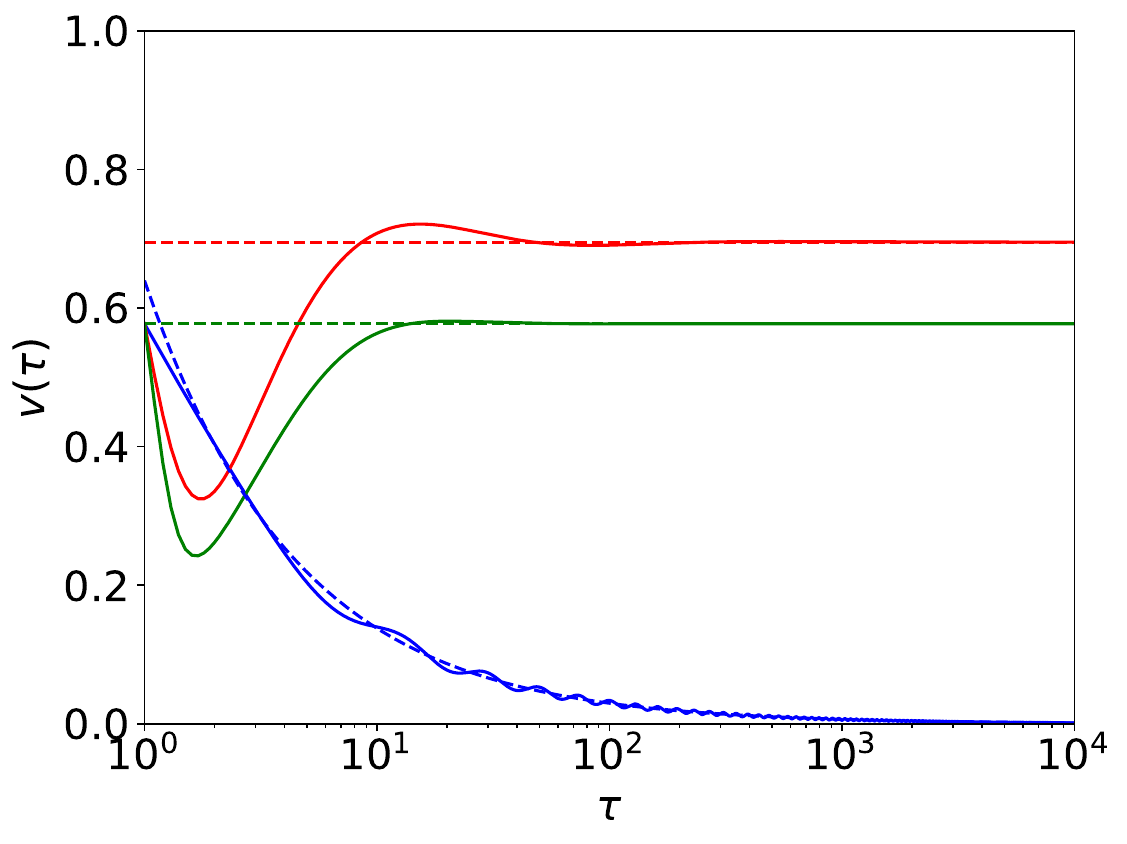}
    \includegraphics[width=0.32\textwidth]{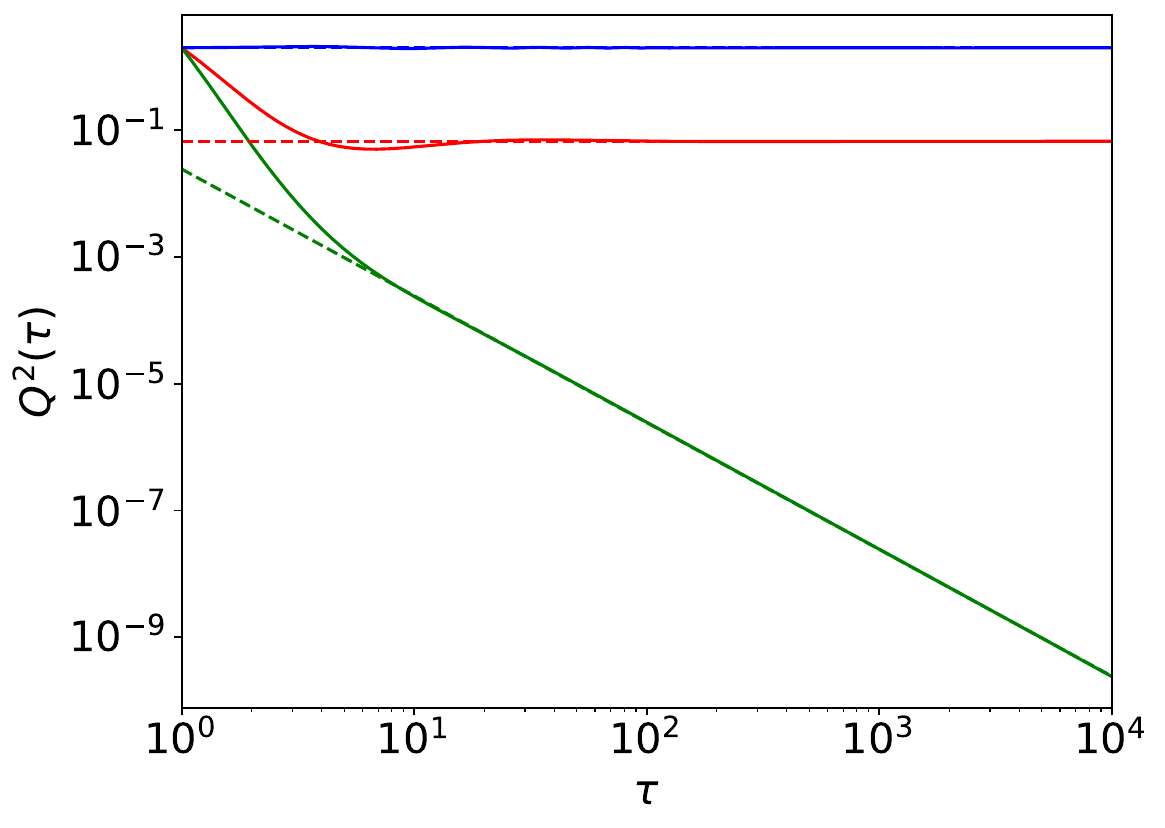}
    \includegraphics[width=0.32\textwidth]{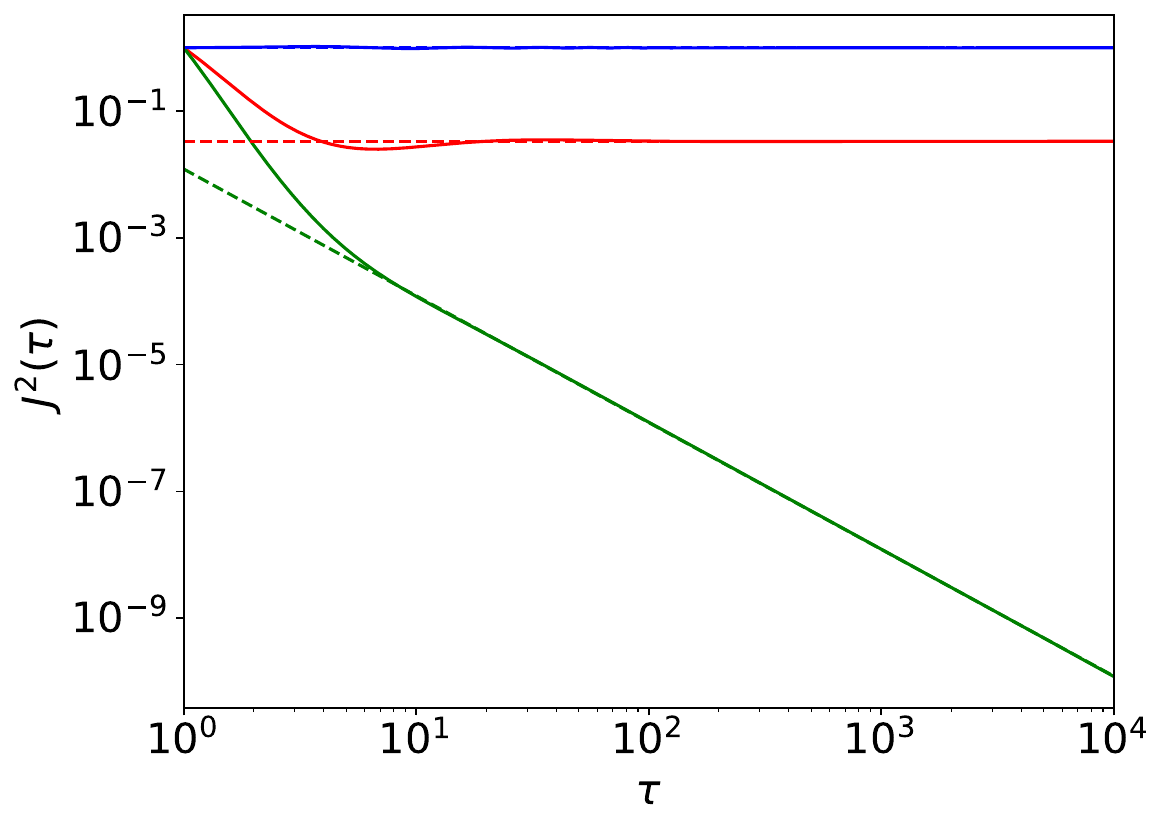}
    \includegraphics[width=0.32\textwidth]{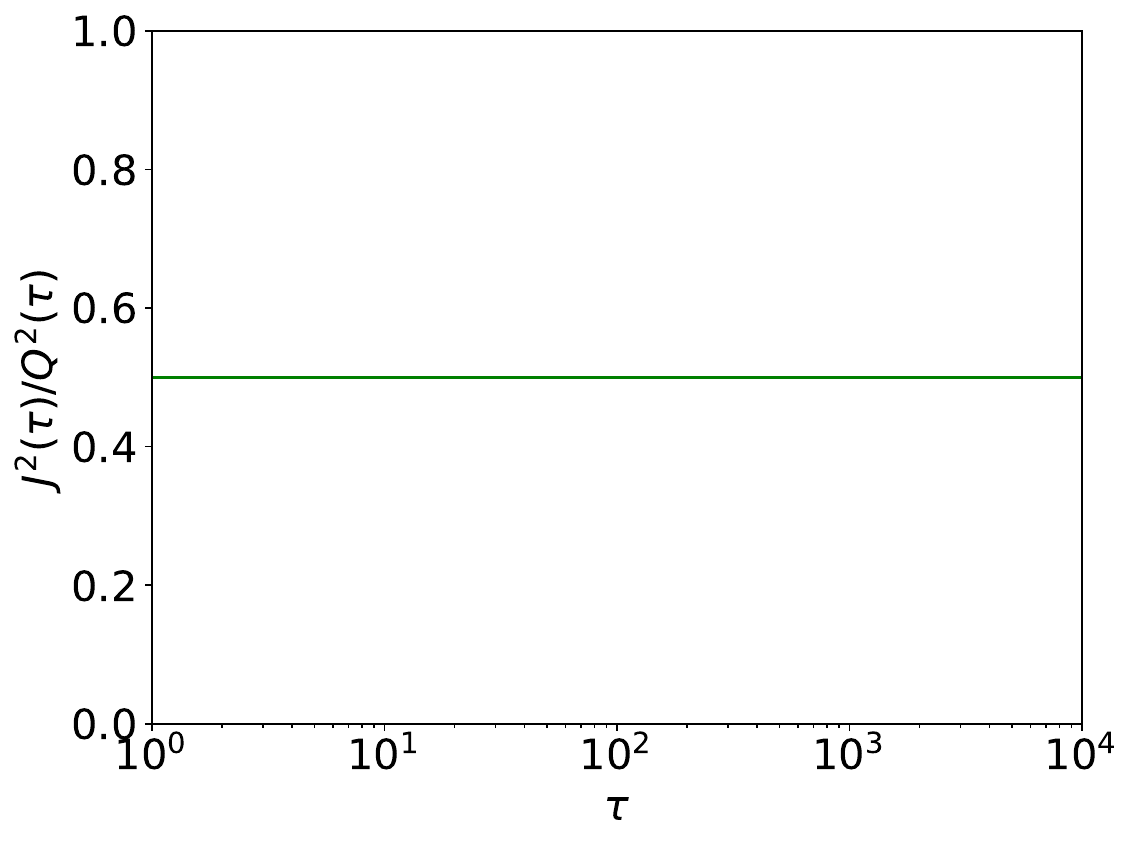}
    \caption{Network evolution as obtained by numerically solving the CVOS equations without loss mechanisms and considering $\dprime{F}\neq0$, for different initial conditions and expansion rates. The three cases correspond to the following initial conditions: (Blue) $\lambda=2/3$, $Q_i=2J_i=0.2$; (Red) $\lambda=2$, $Q_i=2J_i=0.2$; (Green) $\lambda=3$, $Q_i=2J_i=0.2$.}
    \label{fig02}
\end{figure*}

Finally, a constant charge and current solution is also found with arbitrary values of current and charge,
\bsq
\label{solutionA5}
\bq
    L_c &=& L_0\tau\\
    \xi_c&=&\xi_0\tau\\
    J^2 &=& J_0^2\\
    Q^2 &=& Q_0^2\\
    v &=& v_0
\eq
\esq
subject to the constraints
\bsq
\label{constraintsA5}
\bq
    \lambda &=&2\\
    \frac{v_0k_v}{\xi_0}&=&2\\
    v_0^2&=&\frac{1+2\mathcal{K}}{2+2\mathcal{K}}=1-\frac{1}{2+2\mathcal{K}}\,.
\eq
\esq
Here, again, the full scaling solution can only occur in the matter-dominated era. There is no chirality requirement, but there is instead a constraint on the velocity which is set by $\mathcal{K}$, which in turn depends on $\dprime{F}$, and thus we again see the string microphysics impact. Note that the physically expected $0\le v_0^2\le1/2$ corresponds to $-1/2\le\mathcal{K}\le0$.

It is also worthy of note that both of these $\dprime{F}\neq0$ solutions are similar to the ones obtained in \cite{Almeida_2021} and \cite{Oliveira_2012}. Once more, numerical examples starting from arbitrary initial conditions and different expansion rates are presented in Fig. \ref{fig02}. For the lowest expansion rate, the decaying velocity solution from Eq.(\ref{solutionA1}) is obtained.

\section{\label{04losses}Unbiased solutions with energy loss mechanisms}

We now discuss the possible solutions when energy loss mechanisms are allowed, by setting $\tilde{c}\neq0$. Following the earlier discussion, we also set $g=1$; these solutions should also be interpreted as applying to any networks with charge loss mechanisms in which the charge and current decay away. (Nevertheless, we'll present a thorough discussion of the biased solutions in a subsequent work.) Setting $g=1$ also has the consequence of keeping the velocity, charge and current equations the same as before, while the ose for the comoving length scale $L_c$ and correlation length now read
\bsq
\bq
    \alpha &=& \lambda \left[v_0^2\tau^{2\beta}-C_v\mathcal{K}\tau^{2\left(\alpha-\varepsilon\right)+\eta}\right] + \frac{\tilde{c}}{2}\frac{v_0}{\xi_0} \tau^{1+\beta-\varepsilon} \\
    \varepsilon &=& \lambda v_0^2\tau^{2\beta}\left(1+\mathcal{K}\tau^{2\left(\alpha-\varepsilon\right)+\eta}\right)- \nonumber\\
    && -\frac{v_0k_v}{\xi_0}\mathcal{K}\tau^{2\left(\alpha-\varepsilon\right)+\eta+1+\beta-\varepsilon}+\frac{\tilde{c}}{2}\frac{v_0}{\xi_0}\tau^{1+\beta-\varepsilon}
\eq
\esq
One immediately clear point is that imposing $1+\beta-\varepsilon<0$ will reduce the system of equations to the case in the previous section, where no physical solutions were found. This leaves us with the option of $1+\beta=\varepsilon$, for which the CVOS equations become
\bsq
\bq
    \alpha &=& \lambda \left[v_0^2\tau^{2\beta}-C_v\mathcal{K}\tau^{2\left(\alpha-\varepsilon\right)+\eta}\right] + \frac{\tilde{c}}{2}\frac{v_0}{\xi_0}  \\
    \beta &=& \frac{C_vk_v}{v_0\xi_0}\tau^{-2\beta}\left(1+2\mathcal{K}\tau^{2\left(\alpha-\varepsilon\right)+\eta}\right)\nonumber\\
    && -2C_v\lambda \left(1+\mathcal{K}\tau^{2\left(\alpha-\varepsilon\right)+\eta}\right)\\
    \gamma &=& 2\left(\frac{v_0k_v}{\xi_0}-\lambda\right)\\
    \delta &=& 2\frac{\dprime{F}+2J_0^2 \tau^\gamma\ddprime{F}}{\dprime{F}+2Q_0^2 \tau^\delta\ddprime{F}}\left(\frac{v_0k_v}{\xi_0}-\lambda\right) 
    \\
    \varepsilon &=& \lambda v_0^2\tau^{2\beta}\left(1+\mathcal{K}\tau^{2\left(\alpha-\varepsilon\right)+\eta}\right)\nonumber\\
    && -\frac{v_0k_v}{\xi_0}\mathcal{K}\tau^{2\left(\alpha-\varepsilon\right)+\eta}+\frac{\tilde{c}}{2}\frac{v_0}{\xi_0}
\eq
\esq
Moreover, it can be easily seen that decaying velocity solutions are only possible if $\mathcal{K}\tau^{2(\alpha-\varepsilon)+\eta}=-\frac{1}{2}$, which is the same condition as before. This is fully expected, since this constraint is obtained from the velocity equation alone, which is unaffected. Proceeding analogously to the no losses cases, one can conclude that the only possible decaying velocity solution is
\bsq
\label{solutionB1}
\bq
    L_c &=& L_0\tau^{1-\lambda}\\
    \xi_c&=&\xi_0\tau^{1-\lambda}\\
    J^2 &=& J_0^2\\
    Q^2 &=& Q_0^2\\
    v &=& v_0\tau^{-\lambda}\,,
\eq
\esq
subject to the constraints
\bsq
\label{constraintsB1}
\bq
    \mathcal{K}&=&-\frac{1}{2}\\
    \lambda &=& \frac{2}{3}-\frac{\tilde{c}}{3}\frac{v_0}{\xi_0} = \frac{2}{3+\tilde{c}/k_v}\,.
\eq
\esq
This is a generalisation of the solution given by Eqs.(\ref{solutionA1}--\ref{constraintsA1}), to which it reduces for $\tilde{c}=0$. The addition of the energy losses to the damping caused by expansion implies that the single expansion rate for which this solution is allowed becomes slower than the previously found  $\lambda=2/3$. If one wants to express the corresponding expansion rate power law with respect to cosmic time will find:
\begin{equation}
    \lambda_t = \frac{\lambda_\tau}{\lambda_\tau+1} = \frac{2}{5+\tilde{c}/k_v}
\end{equation}
which is the same as the one found by \cite{Oliveira_2012} in the limit $s\to 0$.

Moving to the constant velocity solutions, one finds once more that they can only occur with linear scaling of the characteristic lengths. There are two solutions, which can easily be seen to be the natural extension of Eqs.(\ref{solutionA2}--\ref{constraintsA2}) and Eqs.(\ref{solutionA3}--\ref{constraintsA3}), respectively. The first of these solutions now has the form
\bsq
\label{solutionB2}
\bq
    L_c &=& L_0\tau\\
    \xi_c&=&\xi_0\tau\\
    J^2 &=& J_0^2\tau^{4\lambda v_0^2-2\lambda}\\
    Q^2 &=& Q_0^2\tau^{4\lambda v_0^2-2\lambda}\\
    v &=& v_0\,,
\eq
\esq
subject to the constraints
\bsq
\label{constraintsB2}
\bq
    \dprime{F}&=&0\\
    \lambda &=& \frac{1/v_0^2}{1+\tilde{c}/k_v}\\
    \frac{k_v}{\xi_0}&=&2\lambda v_0 \\
    Q_0^2&=&J_0^2\,,
\eq
\esq
with the caveat that the last constraint is not applicable for $\gamma=\delta=0$. Here one can easily identify a critical velocity which dictates the fate of current and charges at $v_c^2=1/2$; the corresponding critical expansion rate is
\begin{equation}
    \lambda_c = \frac{2}{1+\tilde{c}/k_v}\,;
\end{equation}
again, the energy losses push this critical expansion rate below the matter-dominated era.

\begin{figure*}
    \centering
    \includegraphics[width=0.32\textwidth]{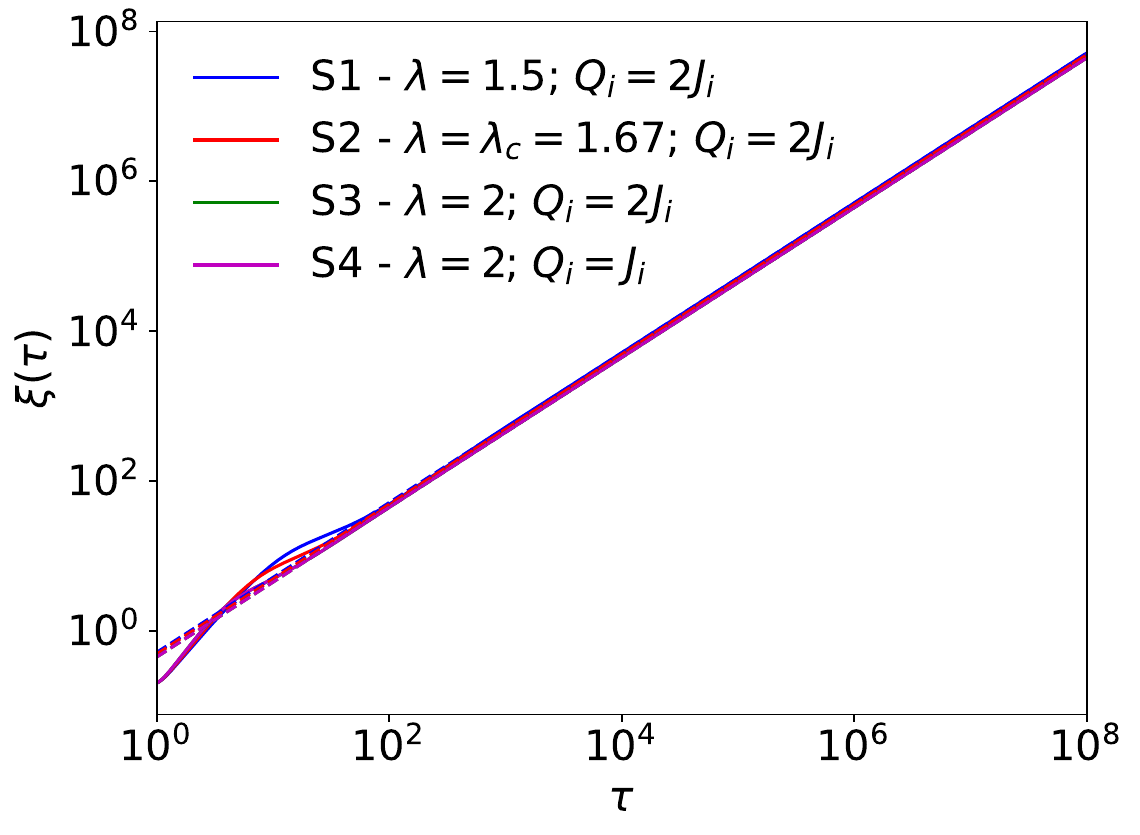}
    \includegraphics[width=0.32\textwidth]{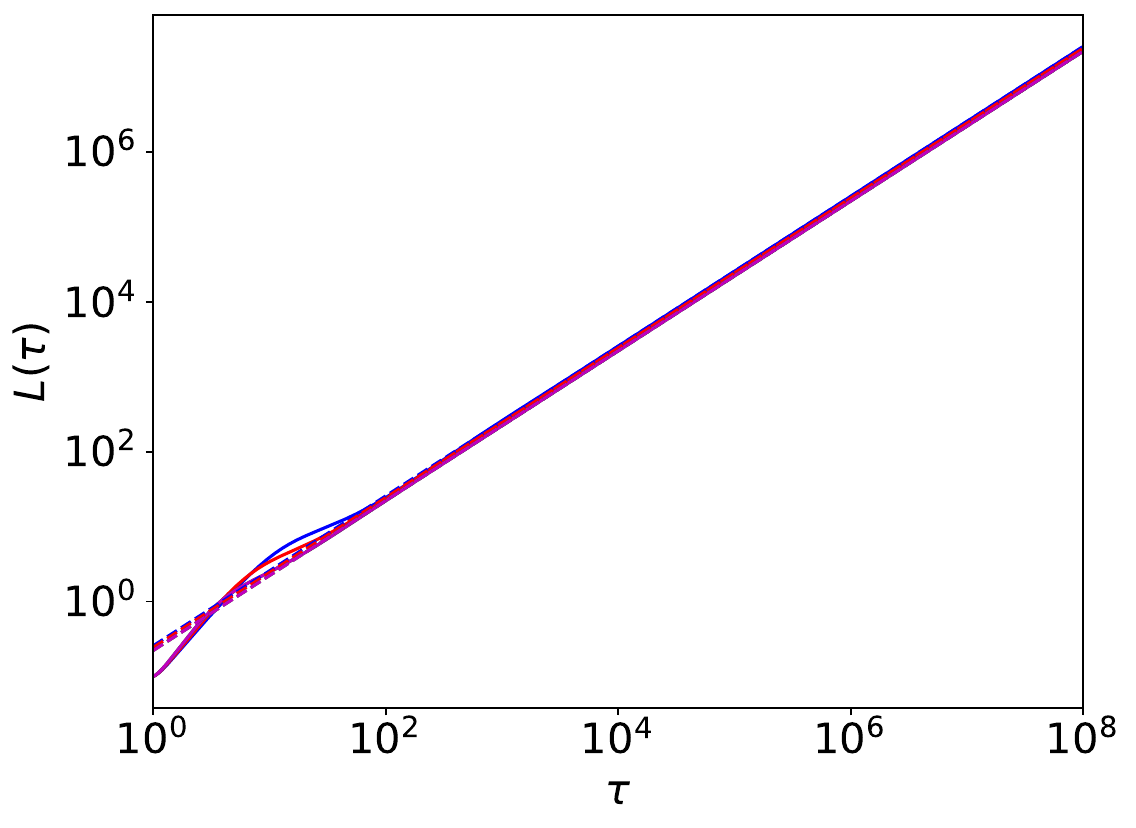}
    \includegraphics[width=0.32\textwidth]{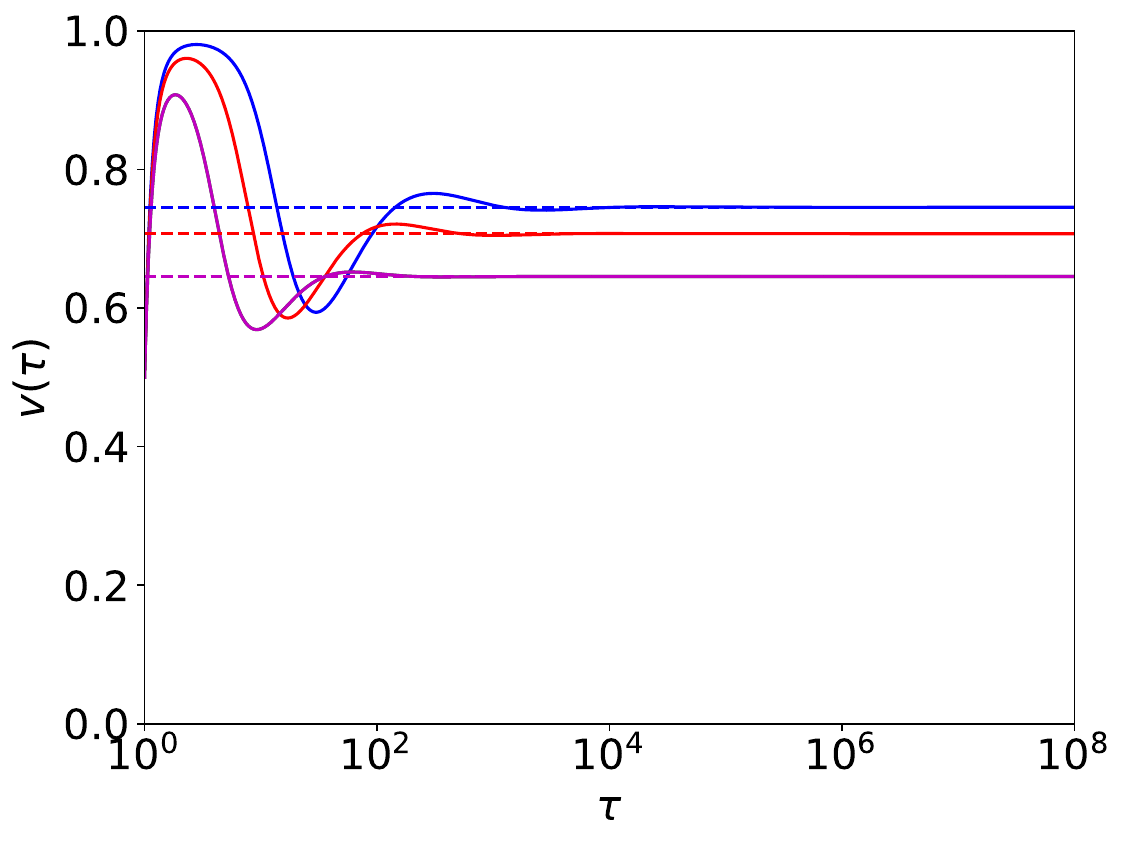}
    \includegraphics[width=0.32\textwidth]{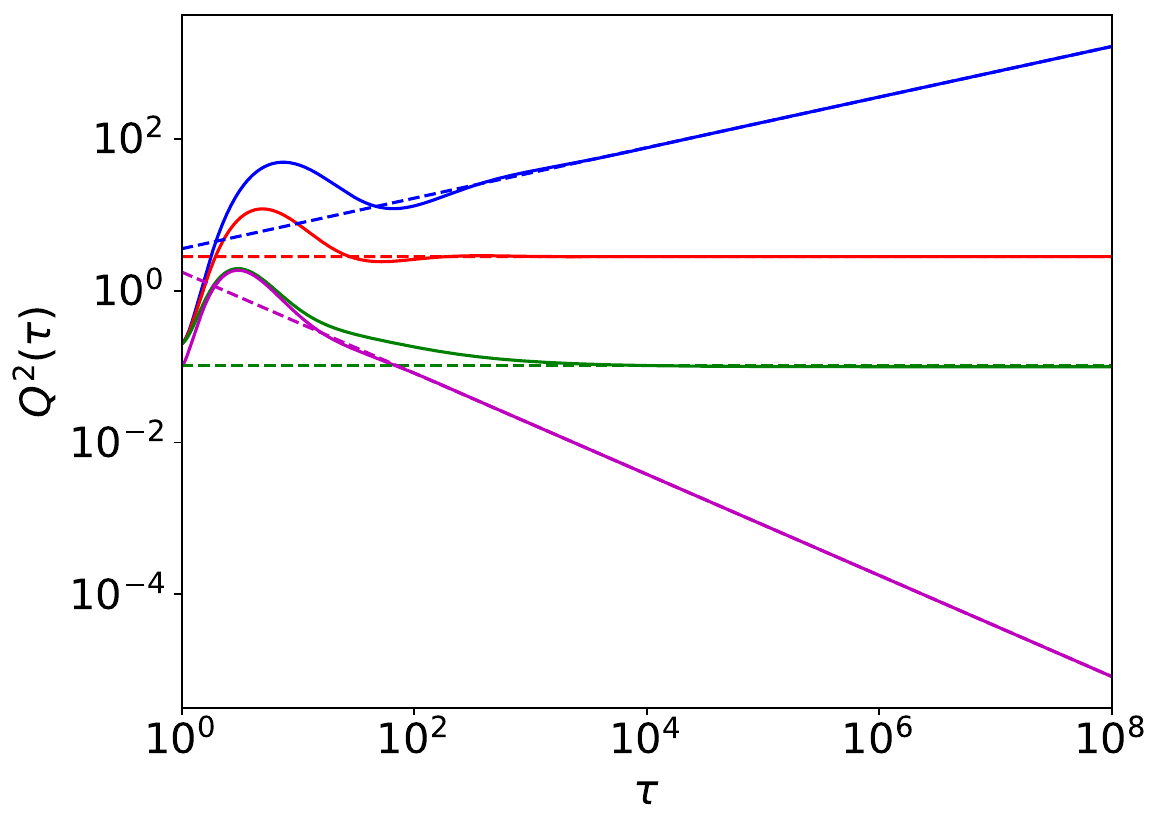}
    \includegraphics[width=0.32\textwidth]{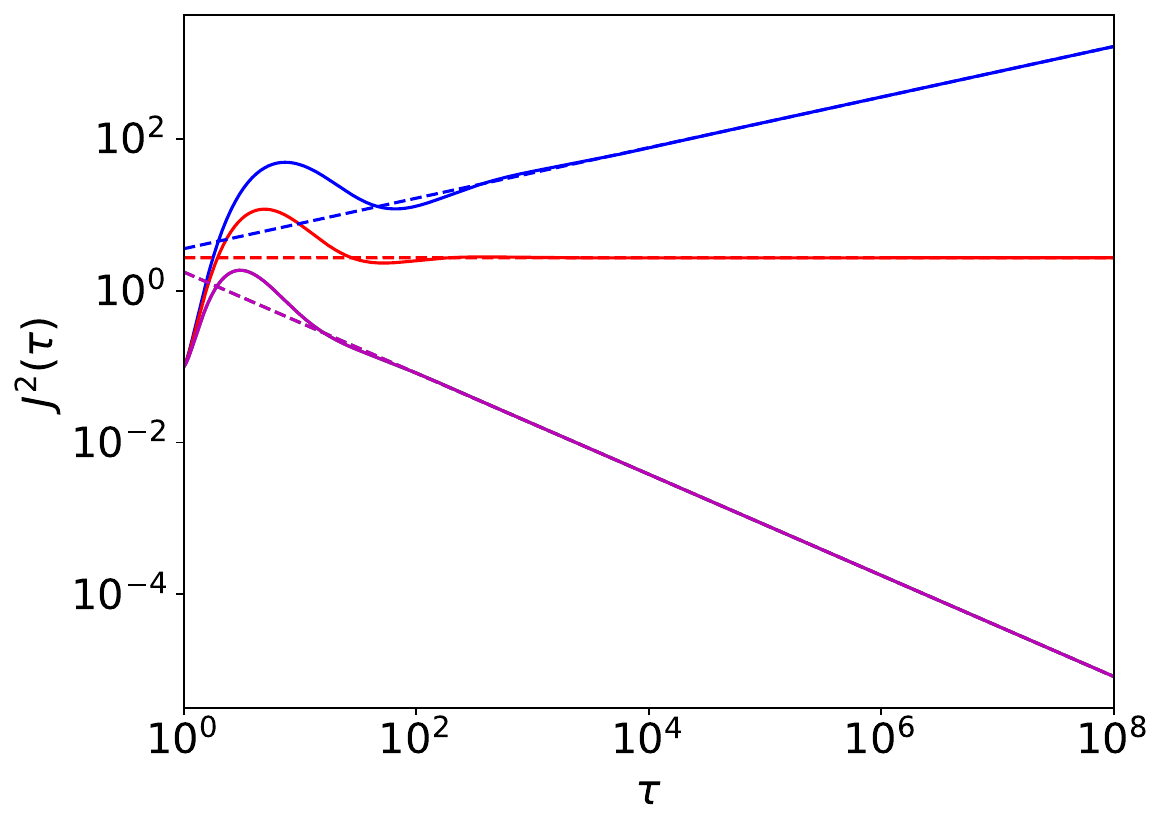}
    \includegraphics[width=0.32\textwidth]{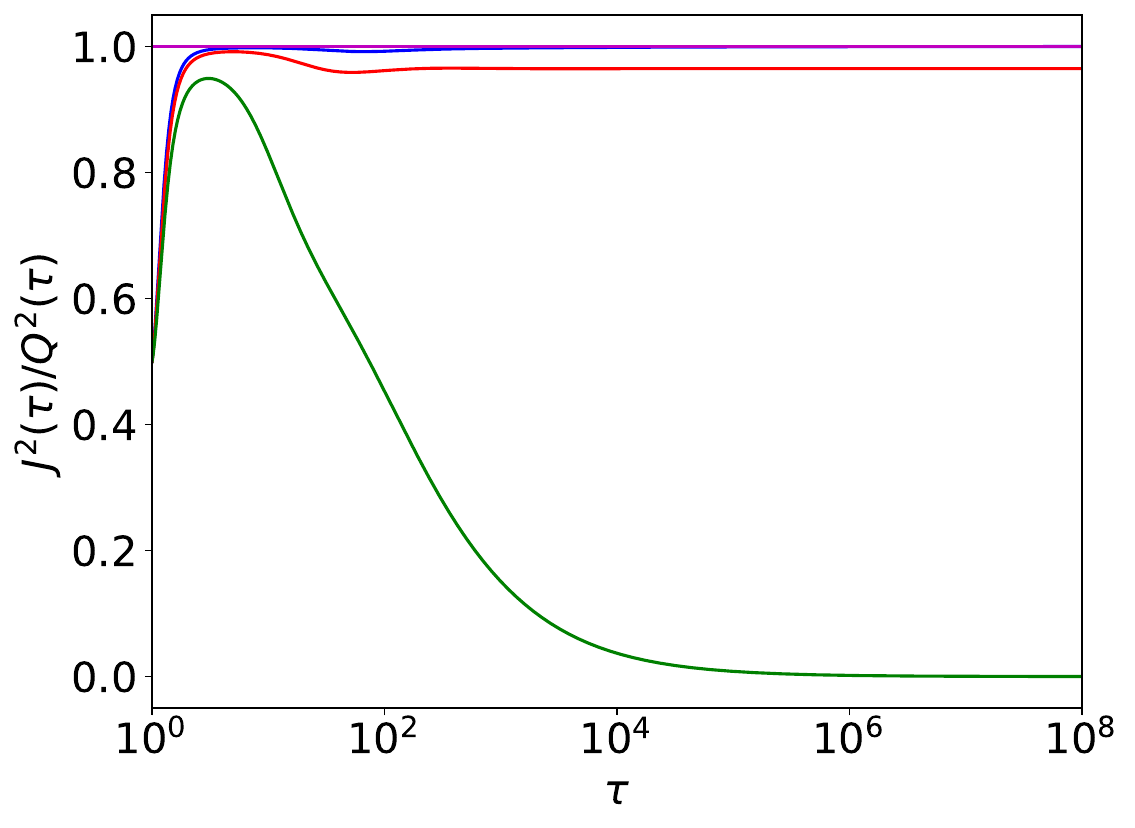}
    \caption{Numerical evolution of networks analogous to those of Figure \ref{fig01} but including energy loss mechanisms, with $\tilde{c}=0.23$. The four cases correspond to the following initial conditions: (Blue) $\lambda=3/2$, $Q_i=2J_i=0.2$; (Red) $\lambda=\lambda_c=1.67$, $Q_i=2J_i=0.2$; (Green) $\lambda=2$, $Q_i=2J_i=0.2$; (Purple) $\lambda=2$, $Q_i=J_i=0.1$.}
    \label{fig03}
\end{figure*}

\begin{figure*}
    \centering
    \includegraphics[width=\columnwidth]{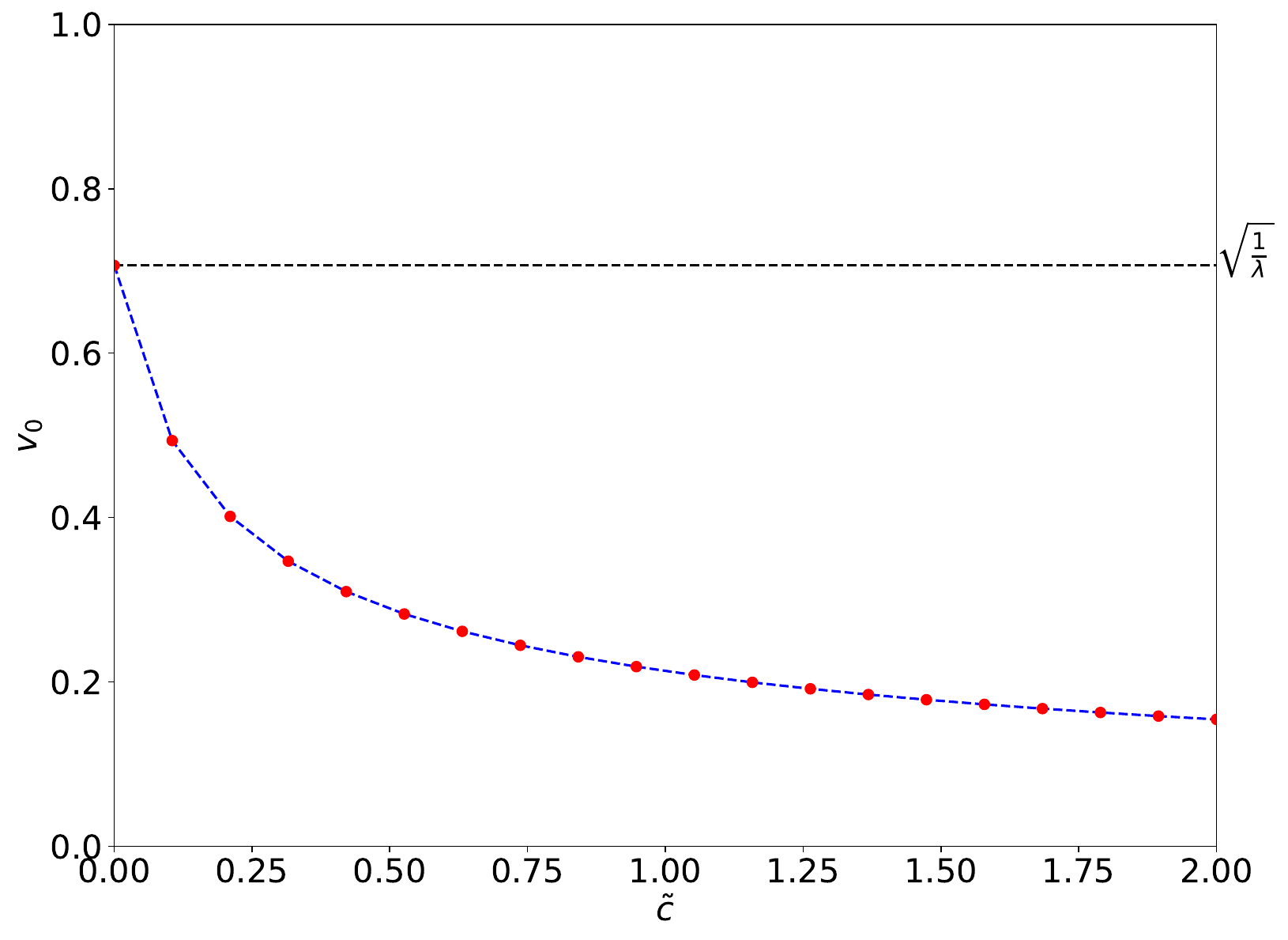}
    \includegraphics[width=\columnwidth]{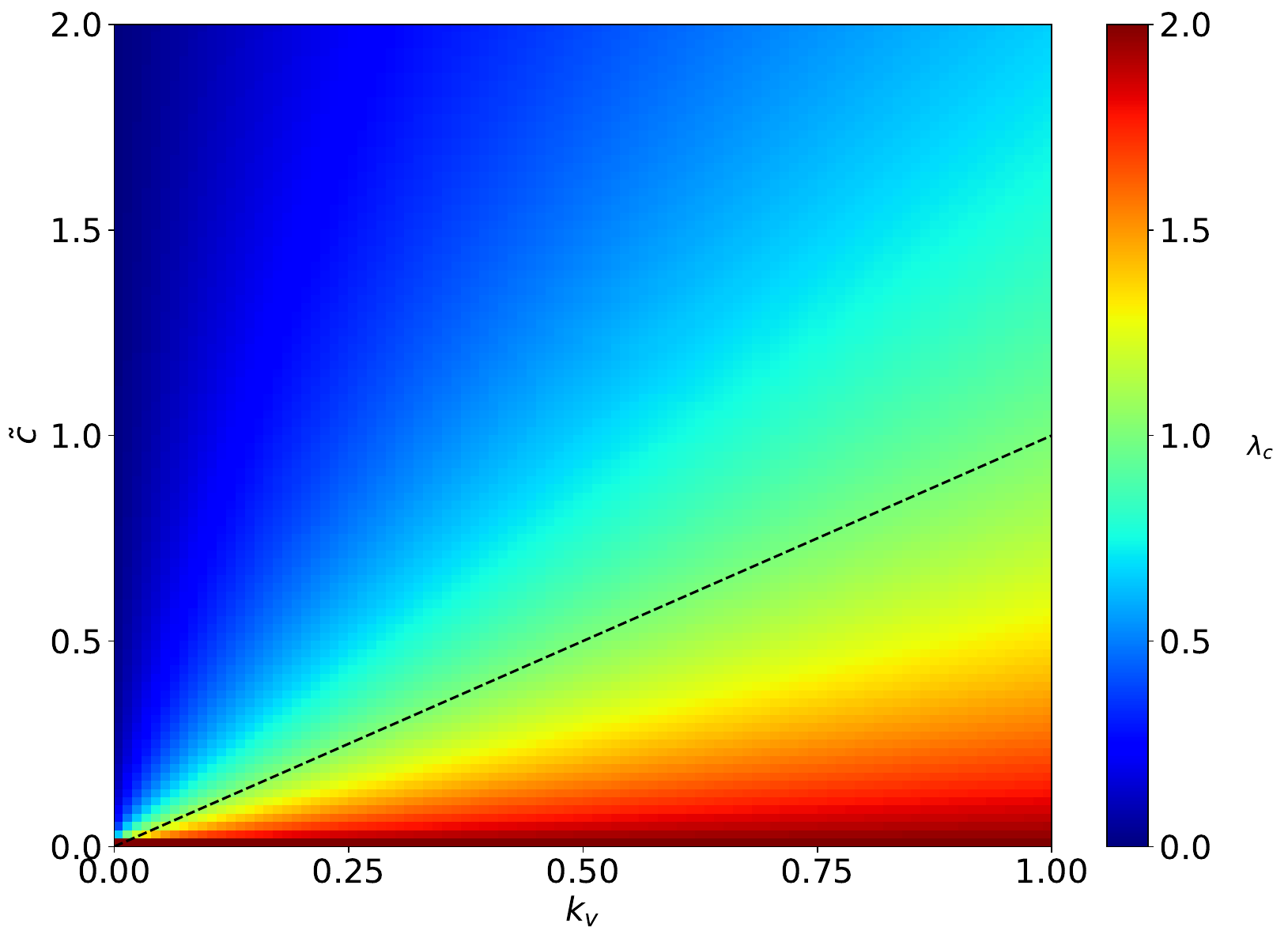}
    \caption{The impact of the energy loss parameter on the asymptotic velocity of the network (left panel) and the critical expansion rate as a function of both the energy loss parameter and the momentum parameter (right).}
    \label{fig04}
\end{figure*}

One the other hand, the constant charge solution is now
\bsq
\label{solutionB3}
\bq
    L_c &=& L_0\tau\\
    \xi_c&=&\xi_0\tau \\
    J^2 &=& J_0^2\tau^{4\lambda v_0^2-2\lambda}\\
    Q^2 &=& Q_0^2 \\
    v &=& v_0\,,
\eq
\esq
subject to the constraints
\bsq
\label{constraintsB3}
\bq
    \dprime{F}&=&0\\
    \lambda &=& \frac{1/v_0^2}{1+\tilde{c}/k_v} > \frac{2}{1+\tilde{c}/k_v}\\
    \frac{k_v}{\xi_0}&=&2\lambda v_0 \\
    v_0^2&<&\frac{1}{2}\,;
\eq
\esq
this is only compatible with decaying currents, or $v_0<v_c$. The relation between the asymptotic velocity and the expansion rate presented in both of these solutions can be inverted to yield
\begin{equation}
    v_0^2 = \frac{k_v}{\lambda\left(k_v + \tilde{c}\right)}\,,
    \label{eq:v_ncl}
\end{equation}
which manifestly reduces to the no loss cases for $\tilde{c}=0$.

As before, a numerical analysis of the evolution governed by the full set of equations is presented in Fig. \ref{fig03}, where in all cases we have used the same initial conditions as in Fig. \ref{fig01}, but we have further assumed that $\tilde{c}=0.23$, a value in agreement with the most recent numerical simulations \cite{Correia_2019}. As was the case for Fig. \ref{fig01}, notice that for too slow expansion rates one would get RMS velocities larger than $1/\sqrt{2}$.

Clearly, for equivalent expansion rates, the asymptotic velocities obtained are lower than in the case of no charge loss (or, in a different perspective, the expansion rate compatible with a given network velocity is lower than before). Figure \ref{fig04} illustrates, for the matter-dominated era ($\lambda=2$), the asymptotic velocities obtained for $\tilde{c}\in[0,2]$ as well as the expected values as computed from Eq.(\ref{eq:v_ncl}) and the no loss limit ($v=1/\sqrt{\lambda}$). Additionally, the critical expansion rate is now slower than the matter era, and solutions in this epoch exhibit the same behaviour of the faster expansion ones from the no loss cases. The impact of the momentum parameter $k_v$ on the value of the critical expansion rate  is also presented in Fig. \ref{fig04}.

If the solutions of Eqs.(\ref{solutionA2}--\ref{constraintsA2}) and (\ref{solutionA3}--\ref{constraintsA3}) have been generalised to Eqs.(\ref{solutionB2}--\ref{constraintsB2}) and (\ref{solutionB3}--\ref{constraintsB3}), the same can be done for  Eqs.(\ref{solutionA4}--\ref{constraintsA4}) and (\ref{solutionA5}--\ref{constraintsA5}). For the first of these, the extended solution is
\bsq
\label{solutionB4}
\bq
    L_c &=& L_0\tau\\
    \xi_c&=&\xi_0\tau \\
    J^2 &=& J_0^2\tau^{(4v_0^2-2)\lambda }\\
    Q^2 &=& Q_0^2\tau^{(4v_0^2-2)\lambda }\\
    v &=& v_0
\eq
\esq
subject to the constraints
\bsq
\label{constraintsB4}
\bq
    \dprime{F}&\neq& 0 \\
    \lambda &=&  \frac{1/v_0^2}{1+\tilde{c}/k_v} > \frac{2}{1+\tilde{c}/k_v}\\
    v_0^2&<&\frac{1}{2}\,,
\eq
\esq
while for the second solution the extended version is
\bsq
\label{solutionB5}
\bq
    L_c &=& L_0\tau\\
    \xi_c&=&\xi_0\tau\\
    J^2 &=& J_0^2\\
    Q^2 &=& Q_0^2 \\
    v &=& v_0\,,
\eq
\esq
subject to the constraints
\bsq
\label{constraintsB5}
\bq
    \lambda &=& 2 - \tilde{c}\frac{v_0}{\xi_0} = \frac{2}{1+\tilde{c}/k_v}\\
    v_0^2 &=& \frac{1+2\mathcal{K}}{2+2\mathcal{K}}\,.
\eq
\esq

\begin{figure*}
    \centering
    \includegraphics[width=0.32\textwidth]{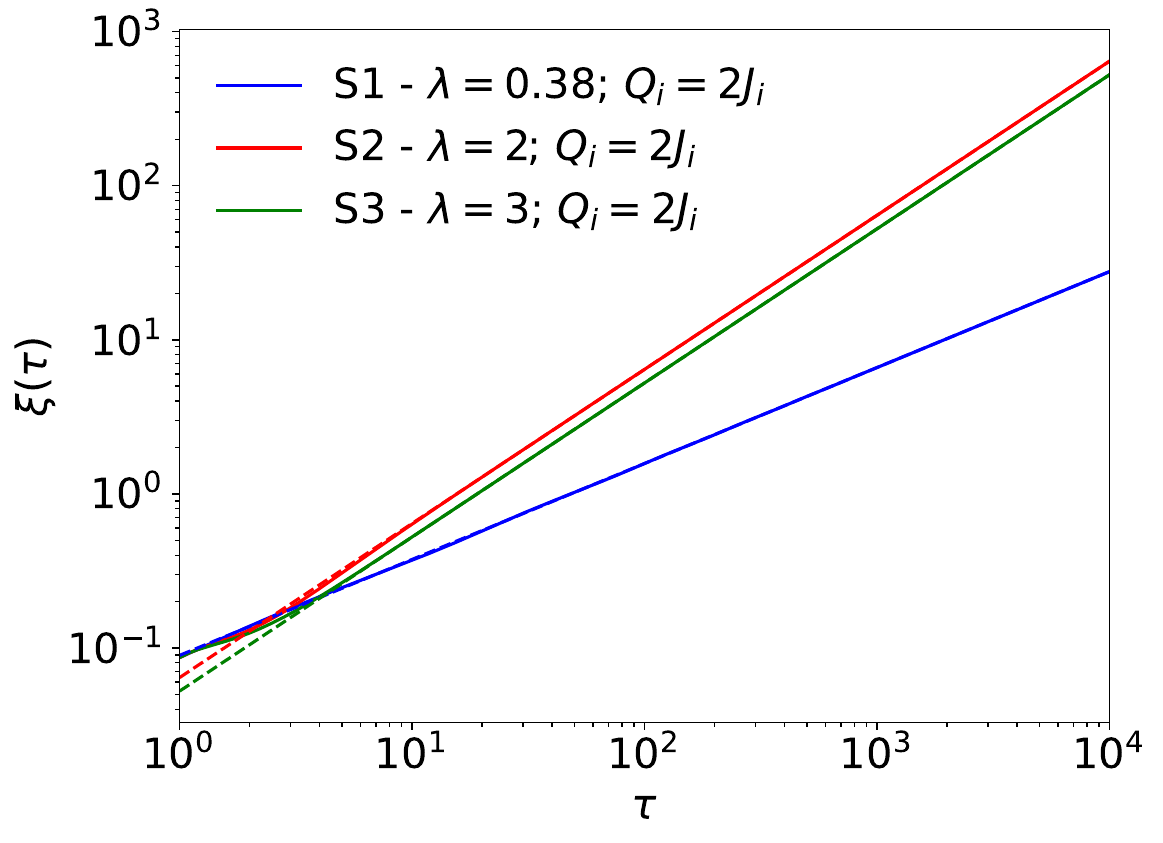}
    \includegraphics[width=0.32\textwidth]{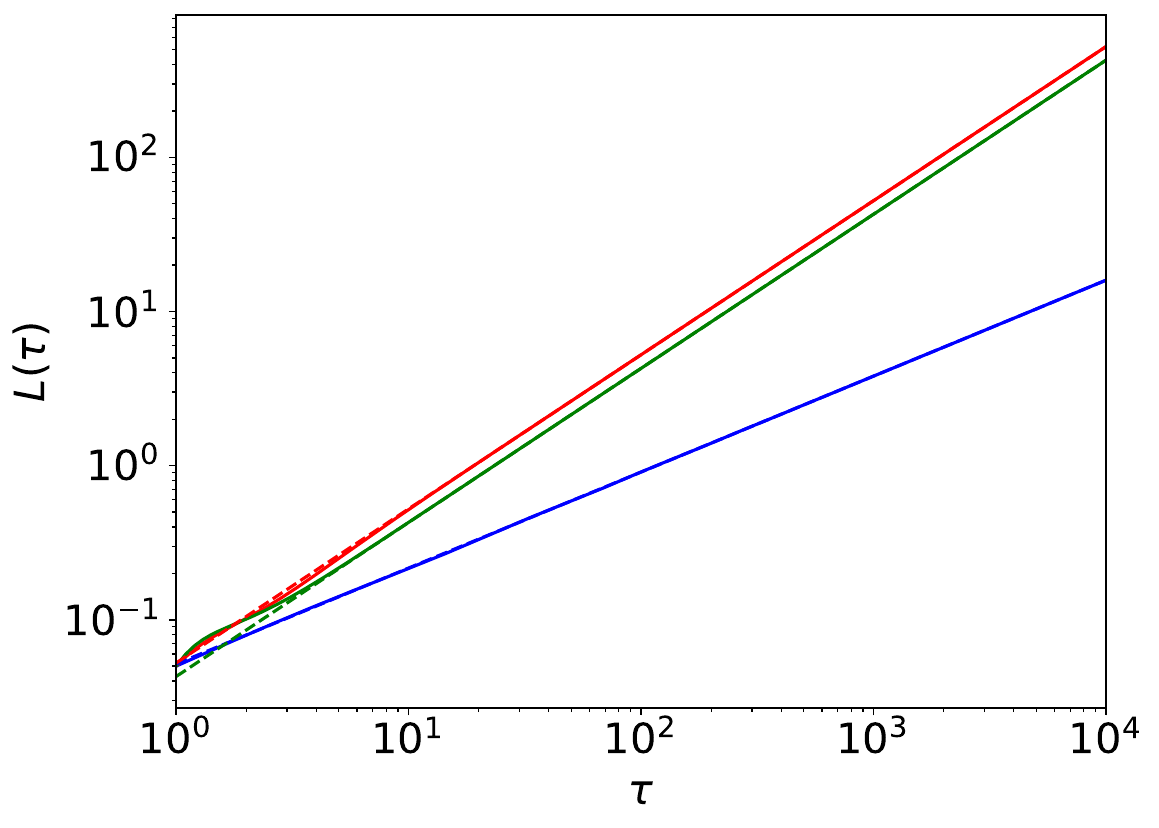}
    \includegraphics[width=0.32\textwidth]{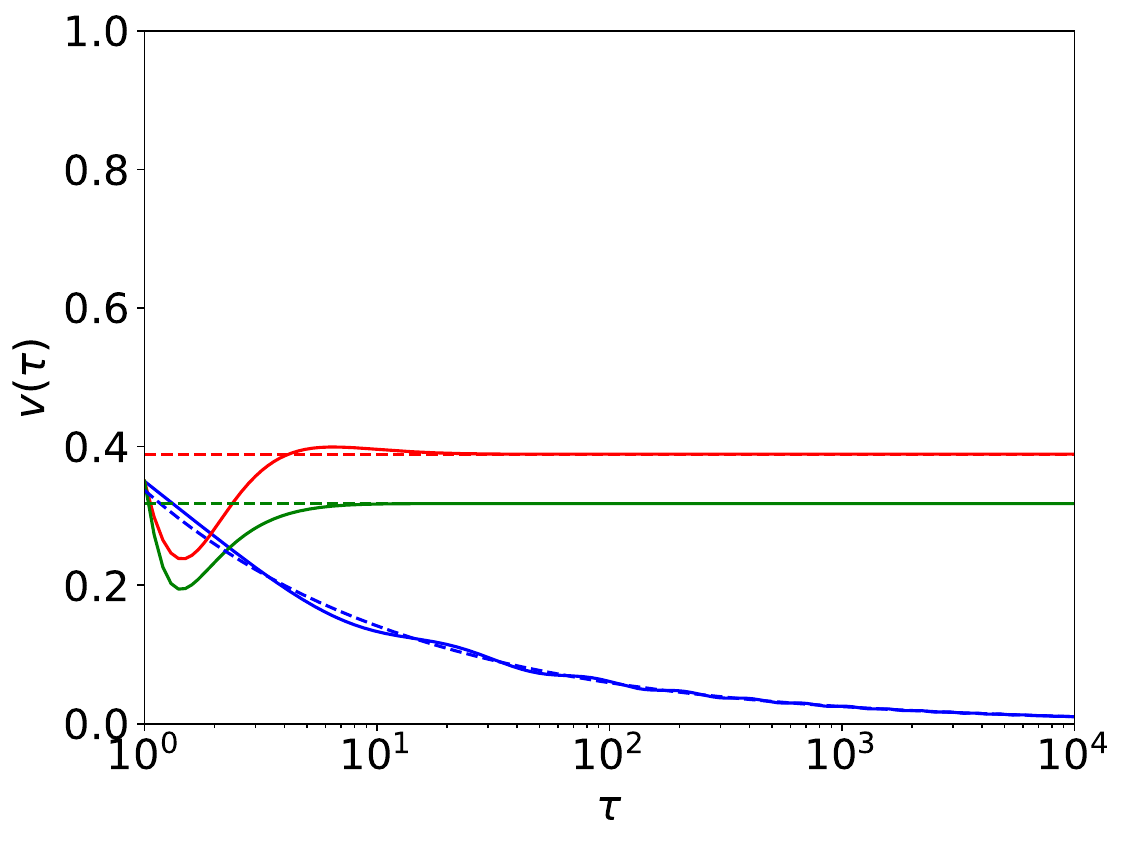}
    \includegraphics[width=0.32\textwidth]{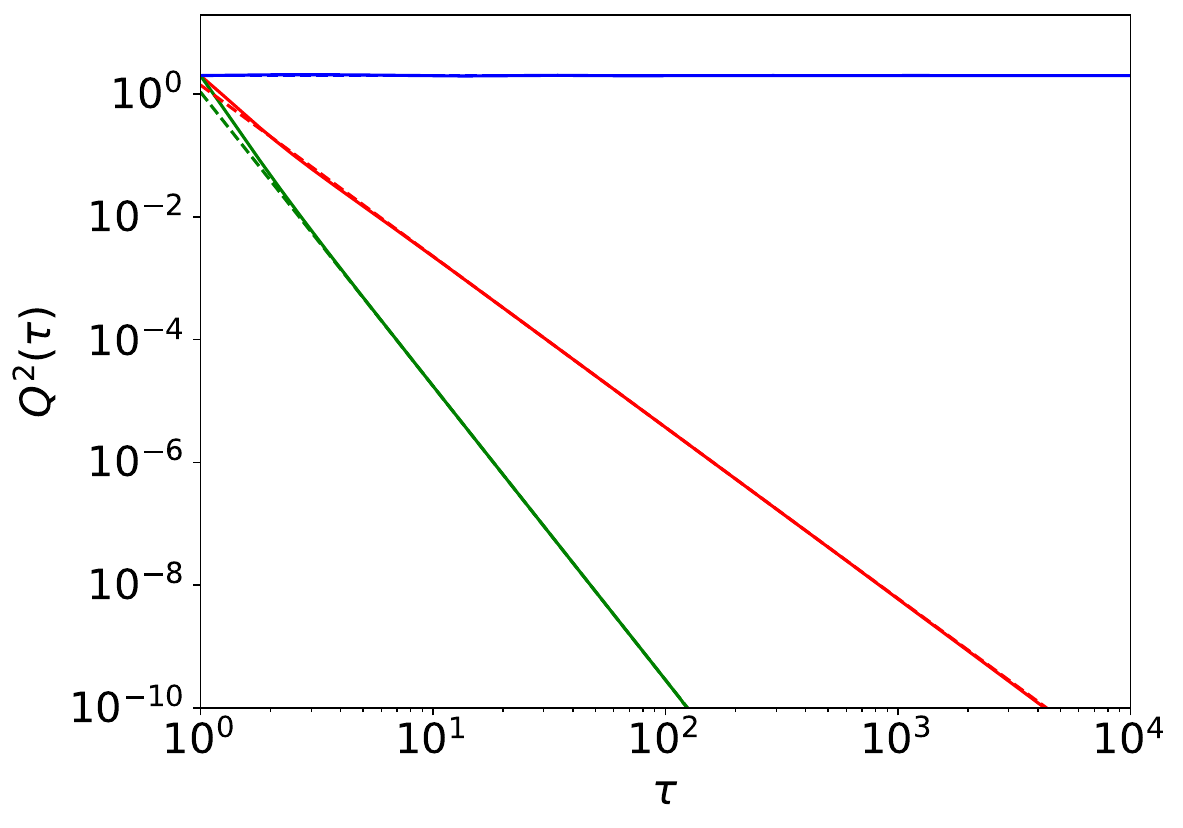}
    \includegraphics[width=0.32\textwidth]{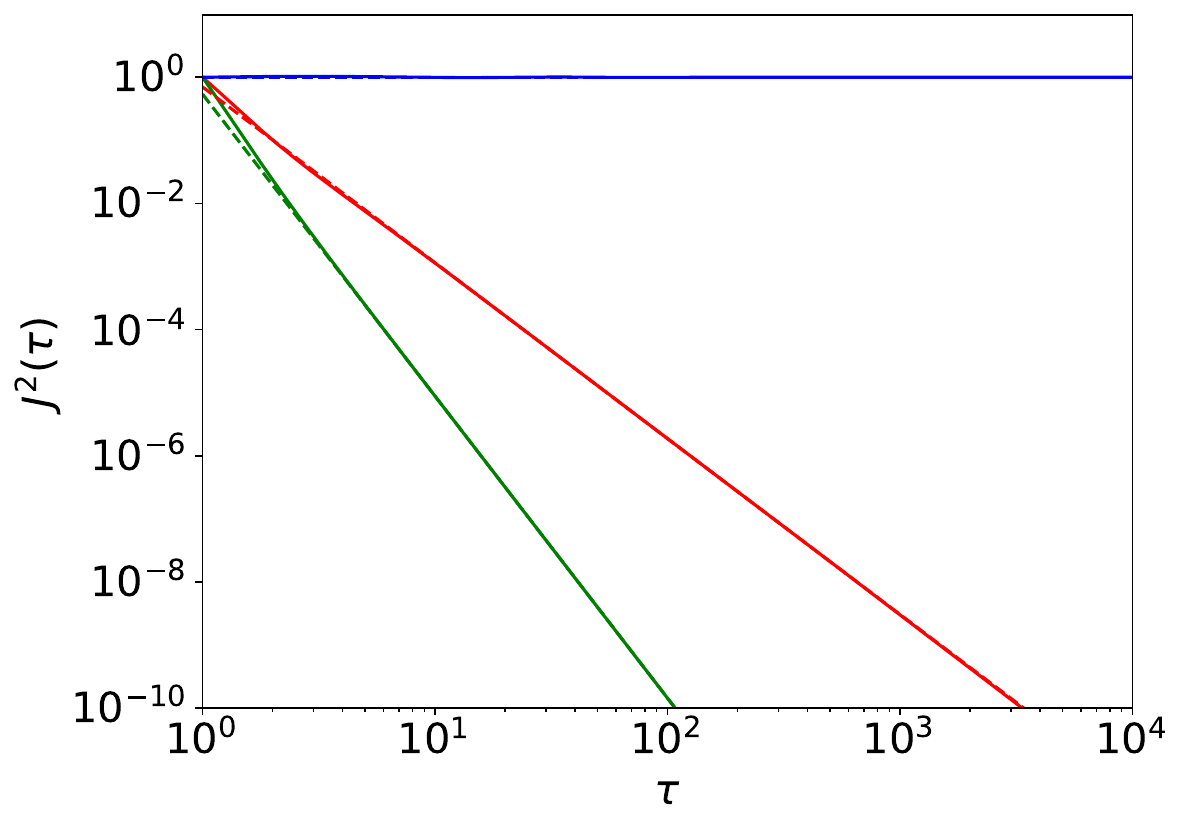}
    \includegraphics[width=0.32\textwidth]{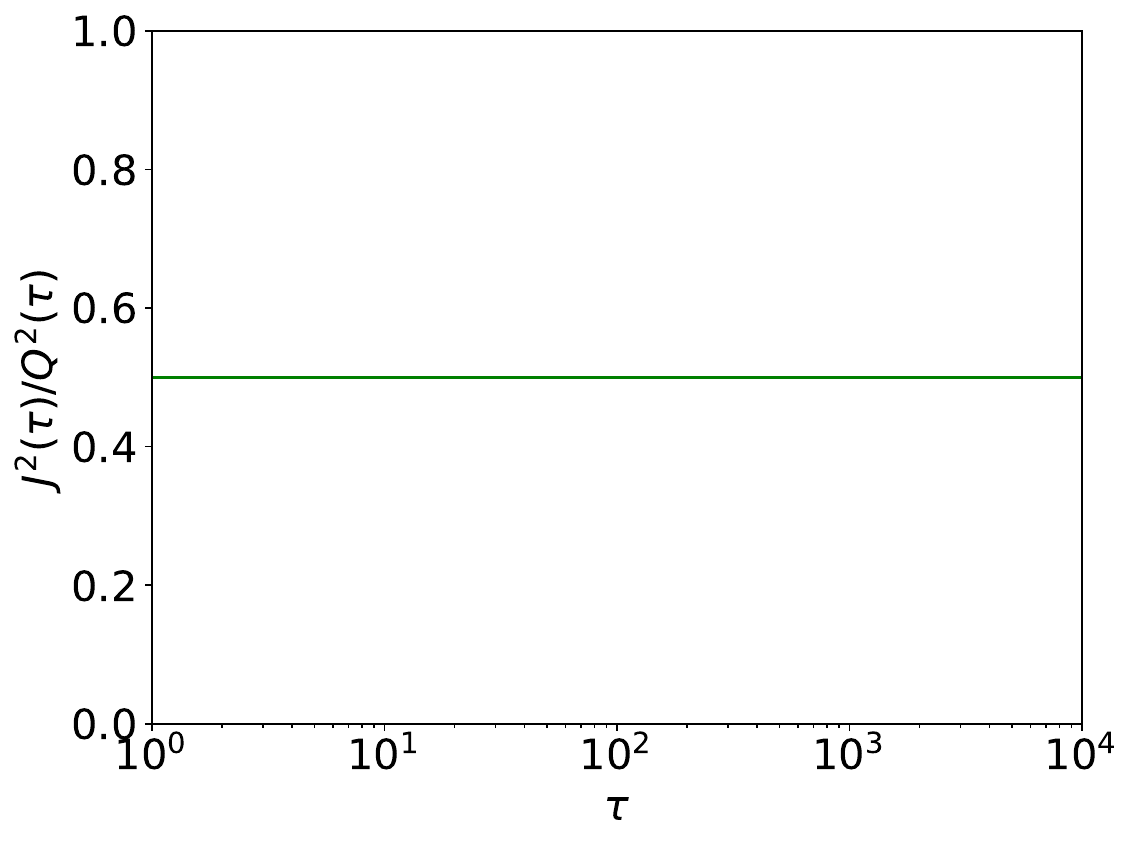}
    \caption{Numerical evolution of networks analogous to those of Figure \ref{fig02} but including energy loss mechanisms, with $\tilde{c}=0.23$. The three cases correspond to the following initial conditions: (Blue) $\lambda=0.38$, $Q_i=2J_i=0.2$; (Red) $\lambda=\lambda_c=2$, $Q_i=2J_i=0.2$; (Green) $\lambda=3$, $Q_i=2J_i=0.2$.}
    \label{fig05}
\end{figure*}

It may may be seen that the critical velocity and expansion rate are still the same. Again, the results of numerical integrations of the CVOS model are shown for different expansion rates in Fig. \ref{fig05} for the cases described by Eqs.(\ref{solutionB1}--\ref{constraintsB1}), (\ref{solutionB4}--\ref{constraintsB4}) and (\ref{solutionB5}--\ref{constraintsB5}), considering the same initial conditions as in the no loss case, and $\tilde{c}=0.23$. By comparison to the results presented in Fig. \ref{fig02}, one can easily see that the asymptotic velocities are now found at lower values.

\section{\label{05conclusions}Discussion and conclusions}

Having discussed the various expanding universe scaling solutions in the previous two sections, and also the Minkowski spacetime ones in Sect. \ref{03nomechanisms}, some general comments and lessons learned are pertinent.

\begin{table*}
    \centering
    \caption{Comparison of the solutions obtained in Sect. \ref{03nolosses} with those obtained by Oliveira {\it et al.} \cite{Oliveira_2012} and Almeida \& Martins \cite{Almeida_2021}, for the chiral and wiggly models, respectively, without energy loss mechanism. The power law exponents have been expressed with respect to conformal time and comoving length scales.} 
    \begin{tabular}{c | c c c c c c | c c c c c | c c c c c }
    \multirow{2}{*}{}  
    & \multicolumn{6}{c|}{Sect. \ref{03nolosses}} 
    & \multicolumn{5}{c|}{Oliveira {\it et al.}  (Chiral)} 
    & \multicolumn{5}{c}{Almeida \& Martins  (Wiggly)} 
    \\
    & $\lambda$ & $\alpha$ & $\beta$ & $\gamma$ & $\delta$ & $\varepsilon$
    & $\lambda$ & $\alpha$ & $\beta$ & $\gamma$ & $\varepsilon$
    & $\lambda$ & $\alpha$ & $\beta$ & $\gamma$ & $\varepsilon$\\
    \hline
   Eqs(\ref{solutionA1}--\ref{constraintsA1})
    & $\dfrac{2}{3}$ & $\dfrac{\lambda}{2}$ & $-\lambda$ & $4-6\lambda$ & $4-6\lambda$ & $1-\lambda$
    & $<2$ & $\dfrac{\lambda}{2}$ & $\dfrac{\lambda-2}{2}$ & $0$ & $\dfrac{\lambda}{2}$
    & $<\dfrac{1}{2}$ & $\dfrac{\lambda}{2}$ & $-\lambda$ &  $2-3\lambda$ & $1-\lambda$
    \\
    \hline
    \multirow{2}{*}{Eqs(\ref{solutionA2}--\ref{constraintsA2})}
    & \multirow{2}{*}{$>1$} & \multirow{2}{*}{1} & \multirow{2}{*}{0} & \multirow{2}{*}{$4-2\lambda$} & \multirow{2}{*}{$4-2\lambda$} & \multirow{2}{*}{1}
    & $2$ & 1 & 0 & 0 & 1
    & $2$ & 1 & 0 & 0 & 1
    \\
    & & & & & & 
    & $>2$ & 1 & 0 & $4-2\lambda$ & 1
    & $>2$ & 1 & 0 & 0 & 1
    \\
     \hline
    Eqs(\ref{solutionA3}--\ref{constraintsA3})
    & $>2$ & 1 & 0 & $4-2\lambda$ & 0 & 1
    & $>2$ & 1 & 0 & $4-2\lambda$ & 1
    & $>2$ & 1 & 0 & 0 & 1
    \\
     \hline
   Eqs(\ref{solutionA4}--\ref{constraintsA4})
    & $>2$ & 1 & 0 & $4-2\lambda$ & $4-2\lambda$ & 1
    & $>2$ & 1 & 0 & $4-2\lambda$ & 1
    & $>2$ & 1 & 0 & 0 & 1
    \\
     \hline
    Eqs(\ref{solutionA5}--\ref{constraintsA5})
    & $2$ & 1 & 0 & 0 & 0  & 1
    & $2$ & 1 & 0 & 0 & 1
    & $2$ & 1 & 0 & 0 & 1
    \\
    \end{tabular}
    \label{tab1}
\end{table*}

When no energy loss mechanism were considered, assuming $\dprime{F}\neq0$ leads to three distinct types of solutions, which can be easily related to the ones identified by \cite{Almeida_2021}, provided one makes the adequate associations between charge and small-scale structure, and also to the ones identified by \cite{Oliveira_2012} for the chiral limit, provided one takes the limit $s\to0$ in these. This relation is detailed in Table \ref{tab1}, where the power law exponents taken from \cite{Oliveira_2012,Almeida_2021} were converted to conformal time and comoving lengths. It is interesting to note that the decaying velocity solution obtained by \cite{Almeida_2021}, although associated with a slightly different expansion rate, shows that the small scale structure plays the role of the charge in our model. In particular, it should be noted that while the small scale parameter, $\mu$ (which physically corresponds to the renormalized mass per unit length of the wiggly strings), evolves as $2-3\lambda$, its square would evolve as $4-6\lambda$, which is exactly the same dependence that we have found.

\begin{table*}
    \centering
    \caption{Comparison of the solutions obtained in Sect. \ref{04losses} with those obtained by Oliveira {\it et al.} \cite{Oliveira_2012} for the chiral limit model, with energy loss mechanism. The power law exponents have been expressed with respect to conformal time and comoving distances.} 
    \begin{tabular}{c | c c c c c c | c c c c c}
    \multirow{2}{*}{}  
    & \multicolumn{6}{c|}{Sect. \ref{04losses}} 
    & \multicolumn{5}{c}{Oliveira {\it et al.} (Chiral)} 
    \\
    & $\lambda$ & $\alpha$ & $\beta$ & $\gamma$ & $\delta$ & $\varepsilon$
    & $\lambda$ & $\alpha$ & $\beta$ & $\gamma$ & $\varepsilon$\\
    \hline
    Eqs.(\ref{solutionB1}--\ref{constraintsB1})
    & $\dfrac{2}{3 + \tilde{c}/k_v}$ & $1-\lambda$ & $-\lambda$ & $0$ & $0$ & $1-\lambda$
    
    & $<\dfrac{2}{1 + c/k_v}$ & $\dfrac{1 + \tilde{c}/k_v}{2}\lambda$ & $\dfrac{1 + \tilde{c}/k_v}{2}\lambda - 1$ & $0$ & $\dfrac{1 + \tilde{c}/k_v}{2}\lambda$
    \\
    \hline
    \multirow{2}{*}{Eqs.(\ref{solutionB2}--\ref{constraintsB2})}   
    & \multirow{2}{*}{$\dfrac{1/v_0^2}{1+\tilde{c}/k_v}$} & \multirow{2}{*}{1} & \multirow{2}{*}{0} & \multirow{2}{*}{$4\lambda v_0^2-2\lambda$} & \multirow{2}{*}{$4\lambda v_0^2-2\lambda$} & \multirow{2}{*}{1}
    & $\dfrac{2}{1 + c/k_v}$ & 1 & 0 & 0 & 1
    \\
    & & & & & & 
    & $>\dfrac{2}{1 + c/k_v}$ & 1 & 0 & $\dfrac{4}{1+c/k_v}-2\lambda$ & 1
    \\
    \hline
     Eqs.(\ref{solutionB3}--\ref{constraintsB3})
    & $>\dfrac{2}{1+\tilde{c}/k_v}$ & 1 & 0 & $4\lambda v_0^2-2\lambda$ & 0 & 1
    & $>\dfrac{2}{1 + c/k_v}$ & 1 & 0 & $\dfrac{4}{1+c/k_v}-2\lambda$ & 1
    \\
    \hline
    Eqs.(\ref{solutionB4}--\ref{constraintsB4})
    & $>\dfrac{2}{1+\tilde{c}/k_v}$ & 1 & 0 & $4\lambda v_0^2-2\lambda$ & $4\lambda v_0^2-2\lambda$ & 1
    & $>\dfrac{2}{1 + c/k_v}$ & 1 & 0 & $\dfrac{4}{1+c/k_v}-2\lambda$ & 1
    \\
    \hline
   Eqs.(\ref{solutionB5}--\ref{constraintsB5})
    & $\dfrac{2}{1+\tilde{c}/k_v}$ & 1 & 0 & 0 & 0  & 1
    & $\dfrac{2}{1 + c/k_v}$ & 1 & 0 & 0 & 1
    \\
    \end{tabular}
    \label{tab2}
\end{table*}

\begin{table*}
    \centering
    \caption{Comparison of the solutions obtained in Sect. \ref{04losses} with those obtained by Almeida \& Martins\cite{Almeida_2021} for the wiggly model, with energy loss mechanism. The power law exponents have been expressed with respect to conformal time and comoving distances.} 
    \resizebox{\textwidth}{!}{
    \begin{tabular}{c | c c c c c c | c c c c c }
    \multirow{2}{*}{}  
    & \multicolumn{6}{c|}{Sect. \ref{04losses}} 
    & \multicolumn{5}{c}{Almeida \& Martins (Wiggly)} 
    \\
    & $\lambda$ & $\alpha$ & $\beta$ & $\gamma$ & $\delta$ & $\varepsilon$
    & $\lambda$ & $\alpha$ & $\beta$ & $\gamma$ & $\varepsilon$\\
    \hline
    Eqs.(\ref{solutionB1}--\ref{constraintsB1})
    & $\dfrac{2}{3 + \tilde{c}/k_v}$ & $1-\lambda$ & $-\lambda$ & $0$ & $0$ & $1-\lambda$

    & $<\dfrac{1}{2+c_{eff}/k_{eff}}$ & $\dfrac{c_{eff}/k_{eff}}{1+c_{eff}/k_{eff}} + \dfrac{1-c_{eff}/k_{eff}}{2+2c_{eff}/k_{eff}}\lambda$ & $-\lambda$ & $\dfrac{2-\left(3+c_{eff}/k_{eff}\right)\lambda}{1+c_{eff}/k_{eff}}$ & $1-\lambda$
    \\
    \hline
    \multirow{2}{*}{Eqs.(\ref{solutionB2}--\ref{constraintsB2})}   
    & \multirow{2}{*}{$\dfrac{1/v_0^2}{1+\tilde{c}/k_v}$} & \multirow{2}{*}{1} & \multirow{2}{*}{0} & \multirow{2}{*}{$4\lambda v_0^2-2\lambda$} & \multirow{2}{*}{$4\lambda v_0^2-2\lambda$} & \multirow{2}{*}{1}
    & $\dfrac{2}{1 + c_{eff}/k_{eff}}$ & 1 & 0 & 0 & 1
    \\
    & & & & & & 
    & $>\dfrac{2}{1 + c/k}$ & 1 & 0 & 0 & 1
    \\
    \hline
    Eqs.(\ref{solutionB3}--\ref{constraintsB3})
    & $>\dfrac{2}{1+\tilde{c}/k_v}$ & 1 & 0 & $4\lambda v_0^2-2\lambda$ & 0 & 1
    & $>\dfrac{2}{1 + c/k}$ & 1 & 0 & 0 & 1
    \\
    \hline
    Eqs.(\ref{solutionB4}--\ref{constraintsB4})
    & $>\dfrac{2}{1+\tilde{c}/k_v}$ & 1 & 0 & $4\lambda v_0^2-2\lambda$ & $4\lambda v_0^2-2\lambda$ & 1
    & $>\dfrac{2}{1 + c/k}$ & 1 & 0 & 0 & 1
    \\
    \hline
    Eqs.(\ref{solutionB5}--\ref{constraintsB5})
    & $\dfrac{2}{1+\tilde{c}/k_v}$ & 1 & 0 & 0 & 0  & 1
    & $\dfrac{2}{1 + c_{eff}/k_{eff}}$ & 1 & 0 & 0 & 1
    \\
    \end{tabular}}
    \label{tab3}
\end{table*}

Remarkably, in all three physical scenarios one finds that the matter-dominated epoch is particularly interesting, since it is the only one allowing for a fully scaling network, with linearly growing characteristic lengths and constant velocity, charge and current. For slower expansion rates the additional degrees of freedom on the string worldsheet impact the dynamics and typically lead to decaying velocities, while for faster expansion rates they typically decay, and the network evolves towards the Nambu-Goto limit.

Our analysis of the full CVOS model also identified a possible fast expansion solution where the current decays but the charge persists, whose physical relevance is not entirely clear. Still, one can find subtle variations of the usual solutions by making particular assumptions on the macroscopic average of the generating function $F$ and its derivative $\dprime{F}$. It is interesting to note that a generalisation of the decaying velocity solutions found by \cite{Oliveira_2012} is only found in the general CVOS for a very particular expansion rate, lower than the radiation epoch one. It should be noted that the universe expansion plays an important role here, since in its absence, and still assuming no additional energy loss mechanism, one would only find frozen network solutions. 

Once the network is allowed to lose energy, even the Minkowski spacetime solutions exhibit some evolution. In that case, and although charge, current and velocity are still constant, the correlation length now grows in time, which is the natural manifestation of a network losing energy. On the other hand, the expanding universe solutions are a generalised version of the ones previously identified. The most distinct features here are the lower asymptotic velocities that come with the increase of $\tilde{c}$ and the change of the critical expansion rate, previously corresponding to the matter epoch, to slower expansion rates. Such a behaviour is fully expected: once additional energy loss mechanisms are available, less damping due to cosmological expansion is needed.

Additionally, it should be noted that while these solutions have been obtained for networks with no charge losses (by setting $g=1$), we expect, based on previous work \cite{Rybak_2023}, that the decaying charge and current solutions are possible solutions for all networks, since these are asymptotically equivalent to a network with $g=1$. Once more, one can easily relate these solutions to the ones obtained by \cite{Oliveira_2012} and \cite{Almeida_2021}, with the appropriate adaptations, as summarised in Tables \ref{tab2} and \ref{tab3} respectively. While the charge decay law of \cite{Oliveira_2012} may not look the same as ours at first sight, referring back to the specific form of the solutions it can be seen that
\begin{equation}
    4\lambda v_0^2 - 2\lambda = \frac{4}{1+\tilde{c}/k_v} - 2\lambda\,.
\end{equation}

Finally, while we leave the full general model, including biases between the bare string and the charge and current, or between the charge and currents themselves, for subsequent work, we can already anticipate the presence of more complex solutions where, for appropriate choices of parameters, the charge and/or current may actually exhibit very distinct behaviours. These solutions will typically be associated with more and more complex constraints on the relations between the different model parameters. For a non expanding universe, however, solutions under these conditions are always associated with constant velocities, but may exhibit constant or growing charges, but not decaying ones, while still requiring $\dprime{F}\neq$0 and $\ddprime{F}=0$.

As has been previously mentioned, a full comparison of the CVOS model prediction with field theory network simulations is not yet possible, although we expect it to be possible in the near future, benefiting from the availability of highly efficient and scalable GPU-based codes \cite{CUDA1,CUDA2}. Such comparisons will confirm the existence (or otherwise) of the scaling solutions predicted by the CVOS model and, moreover, will also provide a possible way of measuring, directly from the simulations, the various phenomenological energy loss parameters, whose values impact the scaling solutions.

One may ask, for example, whether the momentum parameter, which we have assumed to only depend on velocity, as is the case for standard Nambu-Goto strings, should also depend on the charge and current. Another possibility was already hidden in the comparisons with the solutions studied by \cite{Oliveira_2012}, where an additional model parameter, $s$, was used to characterise a possible charge gradient---that possibility has not been considered in the present work, since we have assumed a separable macorscopic model. Comparison with the results of \cite{Oliveira_2012} shows that this assumption does impact the possible solutions; one consequence of this is that this assumption can, at least in principle, be tested in numerical simulations. Moreover, numerically measuring the bias parameters, and their possible dependencies on the charge and current, is particularly important, since one has relatively little \emph{ab initio} physical insight into their possible forms.

Finally, the full generality of the CVOS model is manifest in the fact that the microphysics of the models under consideration, which is encoded in the Lagrangian generating function $f(\kappa)$, is propagated into the function $F$ and its derivatives. As we have shown, some of the possible scaling solutions only occur for specific choices of this function, e.g. $\dprime{F}=0$, so the model predicts (or at least allows) that different current-carrying and superconducting cosmic string networks have substantially different cosmological evolutions. This encoding relies on an averaging process, which in principle can also be tested against numerical simulations. Ultimately, numerically calibrated CVOS models will enable robust predictions for the observational consequences of these networks.


\begin{acknowledgments}
This work was financed by Portuguese funds through FCT (Funda\c c\~ao para a Ci\^encia e a Tecnologia) in the framework of the project 2022.04048.PTDC (Phi in the Sky, DOI 10.54499/2022.04048.PTDC). CJM also acknowledges FCT and POCH/FSE (EC) support through Investigador FCT Contract 2021.01214.CEECIND/CP1658/CT0001 (DOI 10.54499/2021.01214.CEECIND/CP1658/CT0001). 

We gratefully acknowledge the support of NVIDIA Corporation with the donation of the Quadro P5000 GPU used for this research.
\end{acknowledgments}
 
\bibliography{article}

\end{document}